\documentclass[
twocolumn,
prb,preprintnumbers,amsmath,amssymb]{revtex4}

\usepackage{times}
\usepackage{physics}
\usepackage{graphicx}
\usepackage{amssymb}
\usepackage{amsthm}
\usepackage{amsmath}
\usepackage{dsfont}
\usepackage{bm}
\usepackage{mathrsfs}
\usepackage{bbold}
\usepackage{color}

\begin{document}
I
\title{Van der Waals Anomaly}
\author{Itai Y. Efrat and Ulf Leonhardt}
\affiliation{
\normalsize{
Department of Physics of Complex Systems,
Weizmann Institute of Science, Rehovot 761001, Israel}
}
\date{\today}

\begin{abstract}
In inhomogeneous dielectric media the divergence of the electromagnetic stress is related to the gradients of $\varepsilon$ and $\mu$, which is a consequence of Maxwell's equations. Investigating spherically symmetric media we show that this seemingly universal relationship is violated for electromagnetic vacuum forces such as the generalized van der Waals and Casimir forces. The stress needs to acquire an additional anomalous pressure. The anomaly is a result of renormalization, the need to subtract infinities in the stress for getting a finite, physical force. The anomalous pressure appears in the stress in media like dark energy appears in the energy-momentum tensor in general relativity. We propose and analyse an experiment to probe the van der Waals anomaly with ultracold atoms. The experiment may not only test an unusual phenomenon of quantum forces, but also an analogue of dark energy, shedding light where nothing is known empirically. 
\end{abstract}

\maketitle

\section{Introduction}

Van der Waals forces \cite{DLP,Buhmann,Shahmoon} dominate the microcosm of the nanoworld: they cause electrically neutral atoms and molecules to interact with each other on submicron scales. The force is generated by vacuum fluctuations that electrically polarize these particles. Other vacuum forces \cite{Forces} such as the Casimir force \cite{Casimir} we can understand \cite{Rodriguez} as the net effect of van der Waals interactions between the constituents of dielectric media (taking retardation into account \cite{CasimirPolder}). There they cause stresses in the material described by the electric and the magnetic components $\sigma_E$ and $\sigma_M$ of Abraham's stress tensor \cite{Pita,Burger}.  Differences in these stresses create forces.

Van der Waals and Casimir forces in dielectric media need to be renomalized, for otherwise their energy and stress were infinite. The renormalization is local: it depends on the electric permittivity $\varepsilon$ and the magnetic permeability $\mu$ as functions of space $\bm{r}$. We know this from piece--wise homogeneos materials \cite{DLP} where the renomalizer depends on the $\varepsilon$ and $\mu$ of each piece. For sandwiches of several materials, the renormalization procedure \cite{DLP} does not only give finite results, but its quantitative predictions \cite{DLP} have agreed with high--precision measurements \cite{Levitation,CasimirEquilibrium,Decca}. There is thus empirical evidence for the locality of renormalization. 

The question is: {\it how} local is renormalization? This question arises in inhomogeneous media \cite{SimpsonSurprise} where $\varepsilon(\bm{r})$ and $\mu(\bm{r})$ vary gradually. Suppose we would approximate a given inhomogeneous medium with a sequence of homogeneous pieces, making each piece finer and finer. The renormalized stress on each piece is always finite, but the limit is not: the sequence of stresses diverges \cite{Simpson}. From this follows that, if macroscopic electromagnetism can account for vacuum forces in media at all, the renormalizer must also depend on derivatives of $\varepsilon(\bm{r})$ and $\mu(\bm{r})$. It should be still sufficiently local, it cannot depend on all derivatives of $\varepsilon$ and $\mu$, for otherwise the difference between the bare stress and the renormalizer would vanish, the stress would get lost in renormalization. So how local is renormaization? 

We know from planar inhomogeneous media \cite{Grin1} (where $\varepsilon$ and $\mu$ vary in one direction) that the renormalizing Green function must depend on the derivatives of $\varepsilon$ and $\mu$ up to second order, it is not enough to take the gradients of $\varepsilon$ and $\mu$ into account. Furthermore, second--order locality is not only necessary, but sufficient for getting finite stresses, provided $\varepsilon$ and $\mu$ depend on frequency and tend to unity sufficiently fast for large frequencies \cite{Grin1} --- as is the case for real materials \cite{LL8}. Experimental tests of the results of this second--order renormalization procedure have been proposed \cite{Grin2}, but not carried out yet. 

In this paper, we take the next step and study spherically symmetric media. Like in the planar case \cite{Grin1}, the symmetry preserves the polarizations of the electromagnetic field that would normally get mixed in inhomogeneous media, which considerably simplifies the problem. Spherically piece--wise homogeneous media introduce problems of their own \cite{Milton80,Brevik82,Milton97,Brevik98,Brevik99,Barton99,Bordag99,Avni,Comment} but here we avoid them by taking $\varepsilon(r)$ and $\mu(r)$ as gradually varying with radius $r$. 

In our paper we use Lifshitz theory \cite{DLP,LL9,Lifshitz,Scheel} to calculate vacuum forces. This is the theory that agrees best with experiments on the Casimir force \cite{Rodriguez,Decca}. Lifshitz theory uses the fluctuation--dissipation theorem \cite{Scheel} to relate the quantum stress of the vacuum to classical electromagnetic Green functions. The renormalization is carried out by subtracting from the total Green function the outgoing part such that only the scattered part remains. The physical picture behind this renormalization procedure is the idea that van der Waals or Casimir forces \cite{DLP} are caused by the scattering of virtual electromagnetic waves at the boundaries or inhomogeneities of media. The outgoing Green function depends on the local dielectric environment, on $\varepsilon(\bm{r})$ and $\mu(\bm{r})$, and so renormalization is local.  

Like in the planar case \cite{Grin1}, we postulate second--order locality for the renormalizing Green function. However, we find that we need to subtract not only the zeroth--order Green function $D_0$, but also the first-order $D_1$ to get a finite vacuum stress. The Green function $D_0$ describes a purely outgoing wave, whereas $D_1$ captures the first scattering at inhomogeneities around the point of emission. As renormalization is local, the choice of the renormalizing Green function must not depend on the global symmetry of the material --- whether it is planar or spherically symmetric must not matter there. We thus conclude that one should always subtract both the outgoing wave and first--order scattering in renormalization. 

It is not entirely clear what the meaning of this modified renormalization is. Subtracting the outgoing wave has a definite physical meaning: what is left are the scattered virtual electromagnetic waves, and those are the ones causing vacuum forces. But what is the meaning of the first--order scattering we need to subtract as well? Maybe the best description for renormalization is this: in the context of van der Waals and Casimir forces, renormalization is the removal of the near field. Each infinitesimal cell of the medium acts like an emitter and a receiver of virtual electromagnetic waves, but the receiver should not interact with the near field of the emitter. In our procedure we identify and subtract this field. 

In adopting this renormalization procedure, we deduce a remarkable feature of generalized van der Waals forces in inhomogeneous media: we find that the electric and magnetic stresses $\sigma_E$ and $\sigma_M$ should be replaced as
\begin{equation}
\sigma_F \rightarrow \sigma_F +\frac{p}{2} \mathbb{1} 
\label{vdW}
\end{equation}
with $p$ given by Eqs.~(\ref{beta}) and (\ref{pressure}) for spherically symmetric media. This phenomenon we call {\it van der Waals anomaly}, because it corresponds to the trace anomaly of conformally invariant fields in curved space--time \cite{Wald} where the energy--momentum tensor $T^\alpha_\beta$ gets replaced by
\begin{equation}
T^\alpha_\beta \rightarrow T^\alpha_\beta + \varepsilon_\Lambda \delta^\alpha_\beta \,.
\label{wald}
\end{equation}
The energy--momentum contribution $\varepsilon_\Lambda \delta^\alpha_\beta$ has exactly the form of the cosmological constant \cite{Weinberg,HawkingInflation} written on the right--hand side of Einstein's field equations \cite{LL2}, but it is not necessarily constant. Such a contribution to $T^\alpha_\beta$ is called dark energy \cite{Dark}. Following other analogues of cosmology in condensed--matter physics \cite{Volovik,Visser,FF,Kolo1,Kolo2,Kolo3,Stein21} the van der Waals anomaly thus establishes an analogue of dark energy. 

A problem of our modified renormalization procedure is the fact that first scattering, described by $D_1$ in Eq.~(\ref{D1}), is only defined in the asymptotic expansion of the Green function for large frequencies. This expansion captures the infinities of the stress, because those appear in the high--frequency limit, but it does not necessarily describe correctly the finite part of the stress. However, the central quantity of this paper, the anomalous pressure $p$, we calculate to leading order in frequency. Corrections to $p$ will be practically negligible. 

What exactly is the quantity we calculate? As renormalization is local, it may affect identities for the electromagnetic field one would normally take for granted, as they follow from Maxwell's equations, such as 
\begin{equation}
\nabla \cdot (\sigma_E+\sigma_M) = \frac{\nabla\varepsilon}{\varepsilon}\,\mathrm{tr}\,\sigma_E + \frac{\nabla\mu}{\mu}\,\mathrm{tr}\,\sigma_M 
\label{abraham}
\end{equation}
for the electric and magnetic components of Abrahams's stress tensor with
\begin{equation}
\sigma_E = \bm{E} \otimes \bm{D} - \frac{\bm{E}\cdot\bm{D}}{2}\mathbb{1} \,,\quad 
\sigma_M = \bm{B} \otimes \bm{H} - \frac{\bm{B}\cdot\bm{H}}{2}\mathbb{1} 
\label{max}
\end{equation}
where in isotropic dielectric media and in SI units:
\begin{equation}
\bm{D} = \varepsilon_0\varepsilon \bm{E} \,,\quad \bm{B} = \mu_0\mu \bm{H} \,,\quad \varepsilon_0 \mu_0 = c^{-2} \,.
\end{equation}
The Abraham identity of Eq.~(\ref{abraham}) relates the divergence of the stress to gradients in $\varepsilon$ and $\mu$. In our paper we show that the electromagnetic stresses need to be replaced according to Eq.~(\ref{vdW}) for the identity to hold. This is the phenomenon we call van der Waals anomaly. The additional pressure $p$ in the stress we call the anomalous pressure. 

The anomaly is a direct consequence of the locality of our near--field renormalization that takes first scattering into account. In particular, it is the result of the violation of reciprocity by renormalization \cite{Wald}, as follows. Since the renormalizing Green function depends on the local dielectric environment, the emitted and the received Green functions will be different, violating reciprocity. The exchange of virtual electromagnetic waves creates the van der Waals and Casimir forces \cite{DLP,Buhmann}, and so \cite{Case} the recoil imbalance causes a momentum imbalance: the pressure $p$. This anomalous pressure depends only on local dielectric properties, and we have found explicit expressions, Eqs.~(\ref{beta}) and (\ref{pressure}), for the planar and the spherical case. 

Our analysis indicates that the dilute limit of vacuum forces (the limit of vanishing density) is dominated by local dielectric quantities. This is because the van der Waals anomaly dominates the dilute limit. We see this as follows: Eq.~(\ref{abraham}) with replacement (\ref{vdW}) implies that $\nabla\cdot(\sigma_E+\sigma_M)+\nabla p$ vanishes quadratically in density, because $\nabla \varepsilon$ and $\nabla \mu$ go at least linearly with density and the renormalized stresses are also at least linear in density. Now, $p$ depends on the local values of $\varepsilon$ and $\mu$ and their derivatives up to second order, which implies that $p$ is linear in density. We thus conclude that $\nabla\cdot(\sigma_E+\sigma_M)\sim-\nabla p$ in the dilute limit and that the vacuum forces are local. From this also follows that $\nabla\cdot(\sigma_E+\sigma_M)$ is linear in density in the dilute limit \cite{Avni}, in contrast to what has been widely believed \cite{Comment} without questioning. 

The dilute limit opens up the possibility of experimental tests of the van der Waals anomaly with ultracold atoms \cite{Pethick}. In our paper we outline the basic ideas for such tests and estimate the required measurement precision. Note that such experiments would not only probe the anomaly, but also additional assumptions we need to make: we assume that the anomaly gives the electric and the magnetic stress directly and equally, and we assume that the dielectric force is the Helmholtz force \cite{LL8} even for vacuum fluctuations. Our estimation indicates that these measurements are within the range of current experiments.

Such experiments may test more than just a curious phenomenon in Casimir physics, as the van der Waals anomaly is an analogue of dark energy in condensed--matter physics. While dark energy amounts to the most abundant form of mass ($70\%$) in the universe \cite{CMBPlanck}, its theoretical understanding has been riddled with problems \cite{Weinberg}. Testing an analogue of dark energy in the laboratory may give some much--needed empirical guidance to elucidate the matter. 

How exactly is the van der Waals anomaly in dielectric media related to the trace anomaly \cite{Wald} in space--time? Dielectric media appear to electromagnetic fields like curved spaces \cite{Gordon,Plebanski,Schleich,LeoPhil}. In the case of impedance--matched media,
\begin{equation}
\varepsilon=\mu=n \,,
\label{im}
\end{equation}
the effective geometry is exact, with line element
\begin{equation}
ds^2 = n^2 d\bm{r}^2
\label{ds2}
\end{equation}
and, consequently \cite{LL2}, spatial metric tensor $g_{ij}=n^2\mathbb{1}$, its inverse $g^{ij}=n^{-2}\mathbb{1}$ and determinant $g=n^6$. For impedance matching, the electric and magnetic vacuum stresses are the same, $\sigma_E=\sigma_M$, and the total stress is a tensor density \cite{LL2}: $\sigma = \sigma_E+\sigma_M = \sqrt{g}\,\tau$ with $\tau$ being a symmetric tensor $\tau^i_j$. We obtain for the covariant divergence of a symmetric tensor \cite{LL2}:
\begin{eqnarray}
\tau^i_{j;i} &=& \frac{1}{\sqrt{g}}\, \partial_i \sqrt{g}\,\tau^i_j - \frac{1}{2} \left(\partial_j g_{ik}\right)\tau^{ik}
\label{tensordiv} \\
&=& \frac{1}{n^3}\left(\nabla\cdot\sigma - \frac{\nabla n}{n}\,\mathrm{tr}\,\sigma\right) \,,
\end{eqnarray}
which gives Eq.~(\ref{abraham}) in the impedance--matched case of Eq.~(\ref{im}) if we require $\tau^i_{j;i}=0$. Therefore, in impedance--matched media where the medium corresponds to an exact geometry, the Abraham identity is equivalent to the covariant conservation of momentum. In curved space--time, the trace anomaly (\ref{wald}) plays the same role: it serves to restore the covariant conservation of energy--momentum \cite{Wald}. The van der Waals anomaly in dielectric media is thus an analogue of the trace anomaly in space--time, and the trace anomaly appears as dark energy in Einstein's field equations \cite{LL2}.

The van der Waals anomaly might be more than just an analogue, it might actually be the archetype of a mechanism \cite{Annals} that explains dark energy without the need of new physics, as an effect of quantum electromagnetism. This explanation appears to be in agreement with astronomical data \cite{Dror} resolving the tension between the measured Hubble constant and its prediction from the Cosmic Microwave Background \cite{Verde}. It is therefore possible that the same force that rules the microcosm of the nanoworld does also rule the macrocosmos of the expanding universe. Time and tests will tell.

\section{Spherical symmetry}

Our theory is based on the Lifshitz theory \cite{LL9,Lifshitz,Scheel} of vacuum forces in dielectric media \cite{Forces} where renormalization is performed by subtracting the outgoing wave from the bare Green function. The electromagnetic Green function $\mathrm{G}$ describes the vector potential at position $\bm{r}$ created by a point dipole at $\bm{r}_0$ oscillating with frequency $\omega$. It is a bi--tensor with one index referring to the direction of the vector potential and the other to the dipole direction. The vacuum stress is calculated from $\mathrm{G}$ in the limit $\bm{r}_0\sim\bm{r}$ by integration along positive imaginary frequencies and differentiations with respect to $\bm{r}$ and $\bm{r}_0$. If there is a symmetry that preserves the two polarizations of light, the Green bi--tensor gets reduced to two scalar Green functions. In the planar case, these are the Green functions $g_\mathrm{E}$ and $g_\mathrm{M}$ of the electric and the magnetic polarizations. There either the electric or the magnetic field remains orthogonal to the propagation direction and the direction in which the medium varies. In a general inhomogeneous medium, the polarizations mix; it takes a high degree of symmetry to preserve them. 

\subsection{Setting the scene}

Consider a rotationally symmetric medium. We employ spherical coordinates $\{r,\theta,\phi\}$ and require
\begin{equation}
\varepsilon=\varepsilon(r) \,,\quad \mu = \mu(r) \,.
\end{equation}
Without writing this explicitly, we also assume that $\varepsilon$ and $\mu$ depend on frequency (which will become important for renormalization \cite{Grin1}). In spherically--symmetric media, the stress depends only on $r$ and has no shear,
\begin{equation}
\sigma = \mathrm{diag}(\sigma_r^r,\sigma_\theta^\theta,\sigma_\phi^\phi)
\end{equation}
with 
\begin{equation}
\sigma_\theta^\theta = \sigma_\phi^\phi \,.
\end{equation}
The simplest of such media is a homogeneous dielectric sphere embedded in a uniform background \cite{Shell} (with $\varepsilon=\varepsilon_1$, $\mu=\mu_1$ inside and $\varepsilon=\varepsilon_2$, $\mu=\mu_2$ outside the radius of the sphere). Despite extensive effort \cite{Boyer,MDS,Balian,Nesterenko,Milton80,Brevik82,Milton97,Brevik98,Brevik99,Barton99,Bordag99,Avni,Comment} there is still no general solution for the Casimir force in this case. One of the major difficulties of this seemingly simple problem is posed by the sharp interface between the sphere and the background. The issue is that even renormalized vacuum fields tend to infinity at interfaces \cite{LeoBook}. For piece--wise homogeneous planar media these infinities do not matter, but for spherically symmetric media they do. To see this, consider the divergence of the stress (that gives the force density in mechanical equilibrium \cite{Pita}). We get in spherical coordinates \cite{Divergence}
\begin{equation}
(\nabla\cdot\sigma)_r = \partial_r\sigma_r^r + \frac{2}{r}\,\sigma_r^r - \frac{2}{r}\,\sigma_\theta^\theta \,.
\label{div}
\end{equation}
At a spherical interface, $(\nabla\cdot\sigma)_r$ must be a delta function, and so the  radial stress $\sigma_r^r$ must be discontinuous. But both $\sigma_r^r$ and $\sigma_\theta^\theta$ vary and tend to infinity at the interface \cite{Avni}. These diverging contribution cancel each other in the force density \cite{Avni}, but their cancellation is subtle \cite{Avni} and not fully understood when $\varepsilon$ and $\mu$ depend on frequency. For planar media, on the other hand, the stress $\sigma_z^z$ orthogonal to the interface is finite and piece--wise constant, and the diverging stresses in the other directions, $\sigma_x^x=\sigma_y^y$, do not contribute to the force, as $(\nabla\cdot\sigma)_z=\partial_z\sigma_z^z$. There this problem does not occur. Here we avoid it by assuming $\varepsilon(r)$ and $\mu(r)$ to be continuous.

 \subsection{Spectral stresses}
 
Lifshitz theory relates the stress to the electromagnetic Green function in the form of an integral along positive imaginary frequencies \cite{Scheel}:
\begin{equation}
\sigma = \left. - \frac{\hbar c}{2\pi} \int_0^\infty W\,d\kappa \right|_{\bm{r}_0\rightarrow \bm{r}}
\label{stress}
\end{equation}
written here in terms of the imaginary wavenumber $\kappa$ for the frequency $\omega=ic\kappa$. The spectral stresses $W$ depend on the Green function, and on the source and observation points $\bm{r}_0$ and $\bm{r}$. The limit $\bm{r}_0\rightarrow \bm{r}$ is taken in the integral of Eq.~(\ref{stress}) but not before. Spherical symmetry implies that the $W$ are functions of $r$ and $r_0$, and of the angle $\gamma$ between the origin, $\bm{r}$ and $\bm{r}_0$. The limit $\bm{r}_0\rightarrow \bm{r}$ corresponds then to $r_0\rightarrow r$ and $\gamma\rightarrow 0$.

We use the symbol $W_{jF}^{jP}$ to denote the contribution of the $P$-polarization to the $F$--component of the stress in $j$-direction, and obtain (Appendix A):
\begin{eqnarray}
W_{rE}^{rE} &=& \varepsilon \kappa^2 g_E \,,\nonumber\\
W_{rM}^{rE} &=& - \frac{\partial_r r\, \partial_{r_0} r_0\,g_E}{\mu r r_0} - W_{\theta M}^{\theta E} \,,\nonumber\\
W_{\theta M}^{\theta E} &=& \frac{\partial_\gamma \sin\gamma\, \partial_\gamma g_E}{\mu r^2\sin\gamma}
\label{spectralstresses}
\end{eqnarray}
for the electric polarization, and the equivalent expressions for the magnetic polarization with $\varepsilon\leftrightarrow\mu$ and $E\leftrightarrow M$. All other spectral stresses vanish. In the limit of large radii the expressions (\ref{spectralstresses}) reproduce the planar case \cite{Grin1} where $W^{zE}_z=\varepsilon \kappa^2 g_E-\mu^{-1}\nabla\cdot\nabla_0 \,g_E$ for the electric polarization and $W^{zM}_z=\mu \kappa^2 g_M-\varepsilon^{-1}\nabla\cdot\nabla_0 \,g_M$ for the magnetic polarization.  
 
\subsection{Scalar Green functions}

The two polarizations are described by the two scalar Green functions $g_P$. Like in the planar case \cite{Grin1} we write them as
\begin{equation}
g = \nu \nu_0 D
\label{g}
\end{equation}
(dropping the polarization--index) with $\nu_0=\nu(r_0)$ and
\begin{equation}
\nu_E=\mu \,,\quad \nu_M=\varepsilon \,.
\label{nu}
\end{equation}
We show in Appendix A that $D$ obeys the wave equation  
\begin{equation}
\frac{\nabla\cdot\nu\nabla D}{n^2\nu}-\frac{R_r^r}{2} D-\kappa^2 D = \frac{\delta(\bm{r}-\bm{r}_0)}{n^2\nu}
\label{wave}
\end{equation}
in terms of the refractive index $n$ with
\begin{equation}
n^2 = \varepsilon\mu
\label{n2}
\end{equation}
and the abbreviation
\begin{equation}
R_r^r = \frac{2}{n^2\nu}\left(\frac{\nu'^2}{\nu}-\nu''-\frac{\nu'}{\nu}\right)
\label{ricci}
\end{equation}
where the primes denote differentiations with respect to $r$. In the impedance--matched case (\ref{im}) when the medium establishes an exact geometry with line element (\ref{ds2}) the term $n^{-3}\nabla \cdot n \nabla$ in the wave equation (\ref{wave}) is the Laplacian of a scalar and $R_r^r$ becomes the radial component of the Ricci tensor \cite{LL2}. In the planar case \cite{Grin1}, one obtains the same wave equation with $R_r^r$  replaced by $R_z^z$. The wave equations for $D$ thus assume an entirely geometrical form. Finally, for spherical symmetry we have:
\begin{equation}
\frac{\nabla\cdot\nu\nabla}{\nu}=\frac{\partial_r r^2 \nu\, \partial_r}{r^2 \nu} + \frac{\partial_\gamma \sin\gamma\,\partial_\gamma}{r^2\sin\gamma} \,.
\label{laplacian}
\end{equation}
Expressions (\ref{g}-\ref{laplacian}) determine the scalar Green functions and hence the spectral stresses. 

\subsection{Abraham identity}

The Abraham identity, Eq.~(\ref{abraham}), follows from Maxwell's equations in general. Let us see how it follows from the wave equation (\ref{wave}) in our case. We obtain from the expression for $W_{rE}^{rE}$ in Eq.~(\ref{spectralstresses}) by straightforward differentiation
\begin{equation}
\left(\partial_r+\partial_{r_0} + \frac{2}{r}-\frac{\varepsilon'}{\varepsilon}\right)W_{rE}^{rE} = \mathbb{D}_0 D_E
\label{t0}
\end{equation}
in the limit $\bm{r}_0\sim\bm{r}$ where $\mathbb{D}_0$ denotes the differential operator
\begin{equation}
\mathbb{D}_0 = \frac{\kappa^2}{r_0^2}\,\partial_{r_0} r_0 \mu_0\, n^2 + \frac{\kappa^2}{r^2}\,\partial_r r \mu\, n_0^2 \,.
\label{d0}
\end{equation}
We also obtain for $W_{rM}^{rE}$ and $W_{\theta M}^{\theta E}$ in the limit $\bm{r}_0\sim\bm{r}$:
\begin{equation}
\left(\partial_r+\partial_{r_0} + \frac{2}{r}-\frac{\mu'}{\mu}\right) W_{rM}^{rE} - \left(\frac{2}{r}+\frac{2\mu'}{\mu}\right) W_{\theta M}^{\theta E} = \mathbb{D}_1D_E
\label{t1}
\end{equation}
with the differential operator
\begin{eqnarray}
\mathbb{D}_1 &=& \frac{1}{r_0^2}\,\partial_{r_0} r_0 \mu_0\, n^2 \left(-\frac{\nabla\cdot\mu\nabla}{n^2\mu}+\frac{R_r^r}{2}\right) \nonumber\\
&& +\frac{1}{r^2}\,\partial_r r \mu\, n_0^2 \left(-\frac{\nabla_0\cdot\mu_0\nabla_0}{n_0^2\mu_0}+\frac{R_{0r}^r}{2}\right)
\label{d1}
\end{eqnarray}
using Eq.~(\ref{laplacian}) for the Laplacians. For the magnetic polarization we obtain the corresponding expressions by interchanging $E\leftrightarrow M$ and $\varepsilon\leftrightarrow\mu$.

Inspecting the differential operators, we see that the sum $\mathbb{D}_0+\mathbb{D}_1$ applied to $D_E$ contains the left--hand side of the wave equation (\ref{wave}) for $\bm{r}$ and also for $\bm{r}_0$. For $\bm{r}\neq\bm{r}_0$ the right--hand side of Eq.~(\ref{wave}) vanishes. So the sum of the left--hand sides of Eqs.~(\ref{t0}) and (\ref{t1}) vanishes, too, provided $g_E$ satisfies the same wave equation in $\bm{r}_0$ as it does in $\bm{r}$. This is the case if the Green functions are reciprocal,
\begin{equation}
D(\bm{r},\bm{r}_0)=D(\bm{r}_0,\bm{r}) \,.
\end{equation}
The stresses are calculated after the limit $\bm{r}_0\rightarrow\bm{r}$ is taken in Eq.~(\ref{stress}), which implies that $\partial_r\sigma$ corresponds to the sum of $\partial_r W$ and $\partial_{r_0} W$ in the spectral stresses. These and the $2/r$ terms in Eqs.~(\ref{t0}) and (\ref{t1}) form the divergence of the stress according to Eq.~(\ref{div}), and $\sigma_r^r+2\sigma_\theta^\theta=\sigma_r^r+\sigma_\theta^\theta+\sigma_\phi^\phi$ gives $\mathrm{tr}\,\sigma$. We thus obtain the Abraham identity (\ref{abraham}), provided the Green functions are reciprocal. Yet renormalization breaks reciprocity, as we show next.

\section{Renormalization}

The field of Casimir physics began with Casimir's three--page paper \cite{Casimir} in 1948. There he did two things: calculate the force between two perfect mirrors and handwave the justification for renormalization. The latter has haunted the field ever since. Renormalization is necessary, because the bare stress calculated, for example, using the procedure of Sec.~II is infinite. One may view the infinity of the bare vacuum energy and stress as an inevitable feature of quantum field theory or as an artefact of the theory. In either case, the infinity needs to be subtracted to get meaningful results. This process of subtraction is called renormalization. In Lifshitz renormalization \cite{DLP} the subtraction is performed on the Green functions. There the diverging part stems from high frequencies and is therefore well--captured in geometrical optics. 

\subsection{Geometrical optics}

Let us briefly summarize geometrical optics in spherically--symmetric media (further details in Appendix B). In the asymptotic limit of large $\kappa$ the Green function $D$ approaches 
\begin{equation}
D\sim D_0 + D_1 
\end{equation}
with 
\begin{equation}
D_0={\cal A}_0\, e^{-\kappa s}
\label{D0}
\end{equation}
and (for $\bm{r}\sim\bm{r}_0$)
\begin{equation}
D_1=\beta_1 d_1 \,,\quad d_1=\frac{{\cal A}_0 s}{\kappa}\,e^{-\kappa s}
\label{D1}
\end{equation}
where $s$ denotes the optical path length with increment $ds$ given by Eq.~(\ref{ds2}) and ${\cal A}_0$ the amplitude. The contribution $D_0$ describes a purely outgoing wave, while $D_1$ captures the first reflection in inhomogeneities near the point of emission \cite{Grin1}.

Renormalization depends on the local values of $\varepsilon$ and $\mu$. Consider the vicinity of $\bm{r}_0$ up to quadratic order. One verifies that $D_0+D_1$ satisfies the wave equation (\ref{wave}) up to $O(\kappa^{-1})$. One also finds that both $s$ and ${\cal A}_0$ are reciprocal up to quadratic order (Appendix B). From this follows, according to Eqs.~(\ref{D0}) and (\ref{D1}), that $D_0$ and $d_1$ are reciprocal. However, the prefactor $\beta_1$ of $d_1$ turns out to be nonreciprocal: it is given by (Appendix B):
\begin{equation}
\beta_1=\frac{1}{24n^2} \left(\frac{n'^2-2n\nabla^2 n}{n^2} + \frac{6\nu\nu''-9\nu'^2}{\nu^2}\right)
\label{beta}
\end{equation}
evaluated at $\bm{r}=\bm{r}_0$. Equation (\ref{beta}) closely resembles the result \cite{Grin1} for planar media where $\nabla^2n=n''$ and the primes denote $\partial_z$\, whereas here $\nabla^2n=n''+(2/r) n'$ and the primes stand for $\partial_r$. Unlike the planar case \cite{Grin1}, the spherical $D_1$ does contribute to the stress, and this contribution diverges (Appendix C). For getting a finite result one therefore needs to subtract from the bare Green function $D$ both $D_0$ and $D_1$. Appendix D proves the convergence of the stress calculated according to this procedure. As renormalization is local, it must not depend on the global symmetry of the medium --- whether it is planar or spherically symmetric should not play any role. We thus conclude that we should renormalize with $D_0+D_1$ in general, even in the planar case.

\subsection{Reciprocity violation}

The reciprocity--violating first--scattering amplitude $\beta_1$ is going to be responsible for the violation of the Abraham identity (\ref{abraham}). Consider the spectral stresses $W$ in Eq.~(\ref{stress}). Here the violation of the Abraham identity is described by the sum of the left--hand sides of Eqs.~(\ref{t0}) and (\ref{t1}) with definitions (\ref{d0}) and (\ref{d1}). We call this sum $Q$. The wave equation (\ref{wave}) would remove all of the terms in $Q$ if not only $D_0$ were reciprocal, but also $D_1=\beta_1 d_1$. However, in $\nabla\cdot\nu\nabla\,\beta_1(r_0) d_1$ the amplitude $\beta_1$ appears as a constant, whereas in $\nabla_0\cdot\nu_0\nabla_0\,\beta_1(r_0) d_1$ we get the extra terms $d_1\,\nabla_0\cdot\nu_0\nabla_0\,\beta_1 + 2\nu_0(\nabla_0\beta_1)\cdot(\nabla_0 d_1)$. We thus obtain:
\begin{equation}
Q = -\frac{1}{r}\,\partial_r\, r\nu\left(\frac{(\partial_{r_0}\, r_0^2 \nu_0\, \partial_{r_0}\beta_1)}{r_0^2\nu_0} + 2(\partial_{r_0}\beta_1)\partial_{r_0}\right)d_1
\label{q}
\end{equation}
evaluated for $\bm{r}_0\rightarrow\bm{r}$. Note that Eq.~(\ref{q}) is only valid up to $O(\kappa^{-1})$ since $D_0+D_1$ solves the wave equation only up to this order. For large $\kappa$ the dominant term in $Q$ is given by the highest power in $\kappa$. According to Eq.~(\ref{D1}) this term is produced by $\partial_r \partial_{r_0} e^{-\kappa s}$, which is quadratic in $\kappa$ for $\bm{r}_0\rightarrow\bm{r}$. In this limit the optical length $s$ approaches $n\rho$ where $\rho$ denotes the distance $\rho=|\bm{r}-\bm{r}_0|$. Consequently, we obtain:
\begin{equation}
\partial_r \partial_{r_0} e^{-\kappa s} \sim \kappa^2 n^2 e^{-n \rho\, \kappa} \,.
\end{equation}
Furthermore, the amplitude ${\cal A}_0$ approaches $-(4\pi\nu\rho)^{-1}$, because near the point of emission the wave equation (\ref{wave}) reduces to the Laplace equation $\nabla^2 D = \delta(\bm{r}-\bm{r}_0)/\nu$ with this solution. Taking all this into account, we get
\begin{equation}
Q \sim \kappa\,\frac{n^3}{2\pi}\,\beta_1' e^{-n\rho_0\kappa}
\label{qresult}
\end{equation}
where we kept a finite distance $\rho_0$ between emitter and receiver that should be of atomic size. 

\subsection{Anomalous pressure}

Equation~(\ref{qresult}) describes the violation of the Abraham identity for the spectral stress of the renormalizer. We give it a minus sign to account for the subtraction in renormalization, $W=-Q$ in Eq.~(\ref{stress}), and obtain the result
\begin{equation}
(\nabla\cdot\sigma)_r = \frac{\varepsilon'}{\varepsilon}\,\mathrm{tr}\,\sigma_E + \frac{\mu'}{\mu}\,\mathrm{tr}\,\sigma_M - n^3\partial_r \frac{p}{n^3}
\label{abraham1}
\end{equation}
for the total stress $\sigma=\sigma_E+\sigma_M$ with the pressure
\begin{equation}
p=\frac{\hbar c}{(2\pi)^2} \int_0^\infty n^3\left(\beta_{1E}+\beta_{1M}\right) e^{-n\rho_0\kappa} \kappa\,d\kappa \,.
\label{pressure}
\end{equation}
In Eq.~(\ref{abraham1}) we pulled $\varepsilon'/\varepsilon$ and $\mu'/\mu$  out of the spectral integral (\ref{stress}) for simplicity. In reality, $\varepsilon'/\varepsilon$ and $\mu'/\mu$ depend on $\kappa$ and therefore should be left inside the integral or regarded as operators outside. Equation (\ref{pressure}) shows that the anomalous pressure depends on the sum of the scattering amplitudes of the two polarizations. Finally, as $n^3(p/n^3)'=p'-(3/2)(\varepsilon'/\varepsilon+\mu'/\mu)$, $\mathrm{tr}\,\mathbb{1}=3$, and $(\nabla\cdot\sigma)_\theta=(\nabla\cdot\sigma)_\phi=0$, we can cast Eq.~(\ref{abraham1}) in a remarkably simple form: as the van der Waals anomaly (\ref{vdW}) in the Abraham identity (\ref{abraham}).

\subsection{Cosmological considerations}

Without dispersion (frequency dependance of $\varepsilon$ and $\mu$) the anomalous pressure of Eq.~(\ref{pressure}) would diverge like $(n\rho_0)^{-2}$. This does not occur in ordinary dielectric media where $\varepsilon$ and $\mu$ fall off like $\kappa^{-2}$ beyond the last resonance of the material. Therefore the pressure integral (\ref{pressure}) does only logarithmically depend on the atomic size $\rho_0$ (and the precise value of the cutoff $\rho_0$ does not matter in comparison with the other contributions to $p$). But now think of the curved space of general relativity \cite{LL2} as a medium \cite{Gordon,Plebanski,Schleich,LeoPhil}. The equivalence principle \cite{LL2} requires that space acts the same on all objects, in particular on all frequencies of the electromagnetic field --- up to the scale where classical general relativity does no longer hold. It is generally believed \cite{Case} that this scale is given by the Planck length 
\begin{equation}
\ell_\mathrm{P} = \sqrt{\frac{\hbar G}{c^3}} 
\label{planck}
\end{equation}
with $G$ being Newton's gravitational constant. Our analogy would thus suggest that $n\rho_0$ for the medium of space lies in the order of the Planck length $\ell_\mathrm{P}$ of Eq.~(\ref{planck}). In this case, the pressure (\ref{pressure}) may have a cosmologically relevant gravitational effect $Gp$, as both $G$ and $\hbar$ drop out. It depends on the details whether it does. 

According to the cosmological principle \cite{LL2} space is homogeneous and isotropic, as empirically verified in astronomical surveys \cite{Survey}. For  homogeneous and isotropic space, the refractive index is given by \cite{LL2}
\begin{equation}
n = \frac{2n_1}{1+k(r/a)^2}
\label{fish}
\end{equation}
with constant $n_1$ and scale factor $a$ where $a$ gives the radius of curvature and $k\in\{-1,0,1\}$ its sign. Space acts as an impedance--matched medium according to Eq.~(\ref{im}). We obtain from Eqs.~(\ref{beta}) and (\ref{fish}):
\begin{equation}
\beta_{1E} = \beta_{1M} = \frac{1}{6n^3} \left(n''-\frac{n'}{r} - \frac{2n'^2}{n}\right) = 0 \,.
\end{equation}
There is no scattering in maximally symmetric spaces \cite{Fish3D}, and as our anomalous pressure is given by the scattering amplitude, there is no anomaly, provided $a$ is indeed constant.

However, the universe is expanding, the scale factor $a$ is growing with time. The time--dependence of $a$ generates a dynamical Casimir effect \cite{GH} with a trace anomaly \cite{Annals} that also goes as $(n\rho_0)^{-2}$. If we take for $n\rho_0$ the Planck length $\ell_\mathrm{P}$ we obtain the correct order of magnitude of the cosmological constant \cite{Annals}. Moreover, a recent comparison \cite{Dror} with astronomical data has found that the anomaly \cite{Annals} appears to agree not only on the order of magnitude, but with the details of the cosmic expansion, as measured by supernova explosions, Baryon Acoustic Oscillations and the Cosmic Microwave Background. The theory may resolve the Hubble tension \cite{Verde} without invoking new, untested physics \cite{DiValentino} (but rather by revisiting aspects of quantum electromagnetism that have been neglected for decades). The van der Waals anomaly, although not acting in the homogeneous and isotropic space of cosmology, is the archetype of this theory. 

\section{Possible experiment}

Trace anomalies have been theoretically predicted a considerable time ago \cite{Wald}, but they have never been experimentally tested. Here we describe the idea for an experimental test with Bose--Einstein condensates of alkali atoms \cite{Pethick}. At first glance, alkali condensates might not appear as the most likely candidates for demonstrating delicate dielectric effects, as they are dilute gases with low density and hence low dielectric response, but the possibility to optically excite and probe elementary excitations with well--defined spatial structures may very well outweigh this disadvantage.

\subsection{Dilute gases}

Let us first discuss some consequences of the diluteness of Bose Einstein condensates within the context of the van der Waals anomaly that apply to other dilute gases as well. The renormalized vacuum stresses $\sigma_E$ and $\sigma_M$ are caused by the scattering of virtual electromagnetic waves on inhomogeneities. As these are proportional to the number density $\rho$, they too must be at least linear in $\rho$, but so are also $\varepsilon'$ and $\mu'$ in Eq.~(\ref{abraham}). From this and the van der Waals anomaly (\ref{vdW}) follows in the limit of vanishing density:
\begin{equation}
\nabla\cdot(\sigma_E+\sigma_M) \sim - \nabla p \,.
\end{equation}
Assuming that the stress is isotropic in this limit, we integrate:
\begin{equation}
\sigma_E+\sigma_M \sim -p \mathbb{1} \,.
\label{dilutestress}
\end{equation}
As $p$ enters both $\sigma_E$ and $\sigma_M$ equally in the van der Waals anomaly (\ref{vdW}) we further assume that the stresses are equal in the dilute limit, and get
\begin{equation}
\sigma_F \sim -\frac{p}{2} \mathbb{1} \,.
\label{dilutestresses}
\end{equation}
These ideas suggest that in the dilute limit the electric and magnetic stresses are entirely given by the anomalous pressure and hence by local dielectric properties (by $\varepsilon$ and $\mu$ and their derivatives up to second order). 

Consider a dilute gas with $\mu=1$ (such as the alkali Bose--Einstein condensates \cite{Pethick}). In the dilute limit we have:
\begin{equation}
\varepsilon= 1+\chi\,,\quad \chi=\frac{\alpha}{\varepsilon_0}\,\rho
\end{equation}
where $\alpha$ denotes the polarisability \cite{Pethick}. Neglecting in Eq.~(\ref{beta}) all terms of higher order in $\rho$ we obtain in the planar \cite{Grin1} and the spherical case:
\begin{equation}
\beta_{1E}+\beta_{1M} = \frac{\alpha}{6\varepsilon_0}
\begin{cases} \partial_z^2\rho & \quad\mbox{(planar)} \\ 
\left(\partial_r^2-\frac{\partial_r}{r}\right) \rho & \quad\mbox{(spherical).}
\end{cases}
\label{dilutebeta}
\end{equation}
In either case, the resulting van der Waals anomaly is linear in density. This and Eq.~(\ref{dilutestress}) confirms the previous, contraversal result \cite{Avni} that the vacuum stress on a homogeneous sphere in a uniform background is linear in the dilute limit, despite the pairwise nature of van der Waals interactions --- and the very existence of the van der Waals anomaly invalidates the criticism \cite{Comment} as a whole. However, the divergence of the stress does not necessarily give the force density, at least not in mechanical equilibrium \cite{Pita}.

The force in dielectrics is the Helmholtz force obtained by varying the free energy with respect to the density, keeping all external charges and currents constant \cite{LL8}. In this way one gets the force density \cite{LL8}:
\begin{equation}
\bm{f} = \nabla\, \frac{\partial\varepsilon}{\partial\rho}\,\rho\,\frac{\bm{E}\cdot\bm{D}}{2} - \frac{\bm{E}\cdot\bm{D}}{2}\,\frac{\nabla\varepsilon}{\varepsilon} \,.
\end{equation}
In the dilute limit, the Helmholtz force becomes the dipole force \cite{Bjorkholm} with 
\begin{equation}
\bm{f} = -\rho\,\nabla V
\end{equation}
in terms of the dipole potential 
\begin{equation}
V = -\frac{\alpha}{2}\,E^2 = \frac{\alpha}{\varepsilon_0}\,\mathrm{tr}\,\sigma_E \,.
\label{dipole}
\end{equation}
We assume this expression also for vacuum forces (and integrate over the entire spectrum of the polarisability as a function of imaginary wavenumber $\kappa$).

\subsection{Estimation}

Let us first estimate the effect of the van der Waals anomaly on Bose--Einstein condensates. Consider a simple toy model: the particle in a box. Suppose the quantum gas is confined by optical forces to a box of height $a$ (in $z$--direction) and width and length $b$. Assume further that the gas is made non--interacting by Feshbach resonance \cite{Pethick}. One can create excited states in $z$--direction, for example by applying optical pulses \cite{Spectroscopy}. Suppose, for simplicity, that the gas is homogeneous in the other directions. We thus have for the number density
\begin{equation}
\rho=2\rho_0\sin^2k_l z \,,\quad k_l=\frac{\pi}{a}\,l 
\label{rhobox}
\end{equation}
with positive integer $l$, and average density
\begin{equation}
\rho_0=\frac{N}{a b^2} 
\label{rho0}
\end{equation}
where $N$ denotes the total number of atoms. Without atom--atom collisions --- and without van der Waals anomaly --- the energies of these states are just the ones of a quantum particle in a box:
\begin{equation}
E_l = \frac{\hbar^2k_l^2}{2m}
\label{energybox}
\end{equation}
where $m$ denotes the atomic mass. The van der Waals anomaly will create a perturbation of the energies we calculate by first--order perturbation theory, averaging the dipole potential of Eq.~(\ref{dipole}) over the spatial probability distribution of an individual atom $\rho/N$. We obtain from Eqs.~(\ref{pressure}), (\ref{dilutestresses}), (\ref{dilutebeta}) and (\ref{rhobox}-\ref{energybox}):
\begin{equation}
\frac{\langle V\rangle}{E_l} = \delta_0
\label{ratio}
\end{equation}
with
\begin{equation}
\delta_0 = \frac{\rho_0}{4\pi\lambda_C} \int_0^{\kappa_0} \left(\frac{\alpha}{\varepsilon_0}\right)^2 \kappa\,d\kappa 
\label{delta0}
\end{equation}
where $\lambda_C$ denotes the Compton wavelength
\begin{equation}
\lambda_c = \frac{2\pi\hbar}{m c} \,.
\label{compton}
\end{equation}
In the integral (\ref{pressure}) over the imaginary wavenumbers $\kappa$ we have replaced the exponential by a hard cutoff $\kappa_0$. We see that the van der Waals anomaly appears as a density--dependent contribution to the effective mass, independent of quantum number $l$. 

Let us estimate the magnitude of this contribution. The most frequently used atom for Bose--Einstein condensates is Rubidium. It has a Compton wavelength of $\lambda_c\approx 2.5 \times 10^{-18}\mathrm{m}$. As a rough approximation we assume the polarizibility as being constant and of  \cite{Pethick} $\alpha/\varepsilon_0 \approx 300 \times 4\pi a_B^3$ in atomic units (with Bohr radius $a_B$). We assume the cutoff in the optical range: $\kappa_0 = 2\pi/\lambda_0$ with $\lambda_0=0.5\mu\mathrm{m}$. Finally we suppose that $10^6$ atoms are confined in a box of $a=100\mu\mathrm{m}$ and $b=10\mu\mathrm{m}$. This gives $\delta_0=10^{-4}$ as a rough estimate of the relative contribution of the van der Waals anomaly to the energy. The details depend on the actual polarizibilities as a function of imaginary frequency. The effect is small, but not out of the range of precision experiments. 

The modified spectrum can be probed by exciting motional states of the trapped condensate. The excitation lines should vary with density as indicated in Eqs.~(\ref{ratio}) and (\ref{delta0}). The atom--atom collision energy does also vary with density, though. We have assumed that the collisions are reduced to zero by Feshbach resonance \cite{Pethick}, but a residual interaction will always remain in practice. However, its contribution to the energy does depend differently on $l$. As the collision term in the Gross--Pitaevskii equation \cite{Pethick} is proportional to $\rho$, the relative contribution to the energy goes with $\langle\rho\rangle/E_l\propto l^{-2}$ for the density profile of Eq.~(\ref{rhobox}). The different scaling with quantum number $l$ may serve to discriminate between the residual atom--atom collisions and the van der Waals anomaly.

\subsection{Harmonic trap}

Consider now a spherically symmetric harmonic trap of frequency $\Omega$. As before, atomic collisions shall be switched off by Feshbach resonance \cite{Pethick}. Without atom--atom interaction, the spherically--symmetric eigenstates $\psi_l$ obey the Schr\"{o}dinger equation:
\begin{equation}
E_l \, \psi_l = -\frac{\hbar^2}{2m}\left(\partial_r^2+\frac{2}{r}\,\partial_r\right)\psi_l + \frac{m\Omega^2}{2}\,r^2\,\psi_l \,.
\end{equation}
We scale the position $\bm{r}$ in terms of the characteristic length $a$ as
\begin{equation}
\bm{r}=a\bm{\xi} \quad \mbox{with}\quad a^2=\frac{\hbar}{m\Omega} \,,
\end{equation}
and obtain the energies 
\begin{equation}
E_l= \hbar\Omega \left(2l+\frac{3}{2}\right)
\end{equation}
with non--negative integer $l$, and the wave functions
\begin{equation}
\psi_l = \frac{{\cal N}_l}{\xi} \,\exp\left(-\frac{\xi^2}{2}\right)\, H_{2l+1}(\xi)
\end{equation}
in terms of the Hermite polynomials \cite{Erdelyi}. The wave functions are normalized to unity with respect to $\bm{\xi}$ by
\begin{equation}
{\cal N}_l^2= \frac{1}{\pi^{3/2}\, 2^{2l+2}\, (2l+1)!} \,.
\end{equation}
Writing the number density $\rho$ of the condensate as
\begin{equation}
\rho=\rho_0\, \psi_l^2
\end{equation}
with average density
\begin{equation}
\rho_0=\frac{N}{a^3}
\end{equation}
we obtain from Eq.~(\ref{dilutebeta}) in the spherical case the result:
\begin{equation}
\frac{\langle V\rangle}{\hbar\Omega} = \delta_0 \int_0^\infty \left(\partial_\xi^2\psi_l^2 - \frac{\partial_\xi\psi_l^2}{\xi}\right)4\pi\,\xi^2 \,d\xi 
\label{vharmonic0}
\end{equation}
with $\delta_0$ given by Eqs.~(\ref{delta0}) and (\ref{compton}). The integral is dominated by small $\xi$. We apply the asymptotics of the Hermite polynomials \cite{Erdelyi} $H_{2l+1}(\xi)$ for large index and small variable,
\begin{equation}
\psi_l \sim (-1)^l\, \frac{\sin( \sqrt{4l+3}\, \xi)}{(4l+2)^{1/4}\pi \xi} \,,
\end{equation}
and arrive at an approximation of Eq.~(\ref{vharmonic0}) that is quite accurate for $l\ge 1$:
\begin{equation}
\frac{\langle V\rangle}{\hbar\Omega} \sim \frac{\delta_0 }{3\pi^2}\,\frac{(4l+3)^{3/2}}{2l+1} \,.
\label{vharmonic1}
\end{equation}
For $l=0$ we have
\begin{equation}
\frac{\langle V\rangle}{\hbar\Omega} = \frac{\delta_0 }{\sqrt{2}\,\pi^{3/2}} \,.
\end{equation}
In the harmonic trap, the relative contribution $\langle V\rangle/E_l$ of the van der Waals anomaly to the energy varies with $l$ as $(4l+3)^{1/2}/(2l+1)\sim l^{-1/2}$ in contrast to the condensate in a box where $\langle V\rangle/E_l=\mathrm{const}$. This is because harmonic--oscillator states are not rigidly confined, but reach to the classical turning points that depend on the energy, which reduces the density. We can also work out the effect of collisions: the collision term in the Gross--Pitaevskii equation \cite{Pethick} scales with 
\begin{eqnarray}
\langle \rho\ \rangle &\sim& \frac{\rho_0}{2\pi^2}\, \frac{(4l+3)^{1/2}}{2l+1} \quad\mbox{for} \quad l \ge 1 \,,\\
\langle \rho\ \rangle &=& \frac{\rho_0}{(2\pi)^{3/2}} \quad\mbox{for} \quad l = 0 \,. 
\end{eqnarray}
Like for the condensate in a box, the ratio $\langle V\rangle/\langle\rho\rangle$ goes with the energy $E_l$, which allows to discriminate the van der Waals anomaly from residual collisions. 

Overall we found the effect of the anomaly to be in the range of $10^{-4}$ of the energy. The van der Waals anomaly can be distinguished from the influence of residual collisions, and it can be measured by probing the transition frequencies of a trapped Bose--Einstein condensate, following Schawlow's motto for precision measurements \cite{Bres}: ``never measure anything but frequency''.

\section*{Acknowledgements}
We thank 
Y. Avni,
D. Berechya,
N. Davidson,
N. Ebel,
E. Efrati,
I. Griniasty,
and E. Shahmoon for discussions. 
The paper has been supported by the Israel Science Foundation and the Murray B. Koffler Professorial Chair. 

\appendix

\section{Green function and stress}
In this appendix we derive the expression for the Abraham stress as a function of the scalar Green functions $g_E$ and $g_M$. Throughout, we use the vector spherical harmonic notation introduced and discussed in Appendix E. 

\subsection{Scalar Green functions}

We start with $\mathrm{G}(\bm{r}, \bm{r}_0)$, the matrix Green function (the Green bi--tensor) defined as the solution of the inhomogeneous wave equation \cite{LeoBook}:
\begin{equation}
 \label{eq:G}
\nabla\times\frac{1}{\mu}\nabla\times \mathrm{G}+\varepsilon \kappa^2 \mathrm{G}=\mathbb{1}\,\delta(\bm{r}-\bm{r}_0) 
\end{equation}
for imaginary wavelengths $\kappa$. Equation (\ref{eq:G}) is solved by decomposing the matrix Green function into two contributions $\mathrm{G}_E$ and $\mathrm{G}_M$ that originate from the two electromagnetic polarisations preserved in spherically symmetric media, and a contact term $\mathrm{G}_C$: 
\begin{equation}
\mathrm{G} =\mathrm{G}_E+\mathrm{G}_M+\mathrm{G}_C 
\label{eq:polar}
\end{equation}
where each polarisation corresponds to one of the scalar Green functions $g_E$ and $g_M$ as
\begin{eqnarray}
\mathrm{G}_E&=&-\sum_{lm}\frac{g^{lm}_E(r,r_0)}{l(l+1)}\,\bm{\Phi}_{lm}(\theta,\phi)\otimes\bm{\Phi}^*_{lm}(\theta_0,\phi_0)\,,\nonumber\\
\label{eq:GEM}
\mathrm{G}_M&=&\frac{1}{\kappa^2\varepsilon(r)\varepsilon(r_0)}\sum_{lm}\nabla\times \frac{g^{lm}_M(r,r_0)}{l(l+1)}\,\bm{\Phi}_{lm}(\theta,\phi)
\nonumber\\
&&\quad\quad\quad\quad\quad\quad \otimes\,\bm{\Phi}^*_{lm}(\theta_0,\phi_0)\times\overleftarrow{\nabla}_0 
\end{eqnarray}
in terms of the vector spherical harmonics $\bm{\Phi}_{lm}(\theta,\phi)$ described in Appendix E. We are going to specify the contact term later on. Let us first focus on the component $\mathrm{G}_E$ of the $E$--polarization. Consider Eq.~(\ref{eq:G}) applied to $\mathrm{G}_E$ only. From the curl identities of the vector spherical harmonics [Eqs.~(3.12) of Barrera {\it et al.} \cite{Barrera}] follows
\begin{equation}
\nabla\times\frac{1}{\mu}\nabla\times f(r) \bm{\Phi}_{lm} = \frac{1}{r}\partial_r\frac{1}{\mu}\partial_r r f(r)\, \bm{\Phi}_{lm} 
\end{equation}
for an arbitrary function $f(r)$.  We see that for $\mathrm{G}_E$ the left--hand side of the inhomogeneous wave equation (\ref{eq:G}) consists entirely of $\bm{\Phi}_{lm}\otimes\bm{\Phi}^*_{lm}$ terms. We decompose the right--hand side into vector spherical harmonics:
\begin{eqnarray}
 \delta(\bm{r}-\bm{r}_0)\mathbb{1} &=& \frac{\delta(r-r_0)}{r^2}\sum_{lm} \bigg(\bm{Y}_{lm}(\theta,\phi)\otimes\bm{Y}^*_{lm}(\theta_0,\phi_0) \nonumber\\
&&\quad\quad + \frac{\bm{\Phi}_{lm}(\theta,\phi)\otimes\bm{\Phi}^*_{lm}(\theta_0,\phi_0)}{l(l+1)} \nonumber\\
&&\quad\quad + \frac{\bm{\Psi}_{lm}(\theta,\phi)\otimes\bm{\Psi}^*_{lm}(\theta_0,\phi_0)}{l(l+1)} \bigg)
\label{eq:delta}
\end{eqnarray}
and see that Eq.~(\ref{eq:G}) is satisfied for the $\bm{\Phi}_{lm}\otimes\bm{\Phi}^*_{lm}$ terms if we require:
\begin{equation}
\frac{1}{r}\partial_r\frac{1}{\nu}\,\partial_r rg^{lm} -\frac{l(l+1)}{r^2\nu}g^{lm} - \frac{n^2\kappa^2}{\nu}\,g^{lm}=\frac{\delta(r-r_0)}{r^2}
\label{eq:g} 
\end{equation}
dropping the polarization label and using $\nu=\mu$ according to Eq.~(\ref{nu}) and the refractive index $n$ of Eq.~(\ref{n2}). This settles the electric polarization. 

Now turn to the magnetic polarization. It should take care of the remaining  $\bm{Y}_{lm}\otimes\bm{Y}^*_{lm}$ and $\bm{\Psi}_{lm}\otimes\bm{\Psi}^*_{lm}$ terms in Eq.~(\ref{eq:delta}) for $\delta(\bm{r}-\bm{r}_0)\mathbb{1}$. Let us define the contact term
\begin{equation}
\mathrm{G}_C = \frac{\delta(r-r_0)}{\kappa^2 r r_0\, \varepsilon(r_0)} \sum_{lm} \left(\bm{Y}_{lm}\otimes\bm{Y}^*_{lm} + \frac{\bm{\Psi}_{lm}\otimes\bm{\Psi}^*_{lm}}{l(l+1)} \right)
\label{eq:contact} 
\end{equation}
where we omit the dependancies on the various angles for simplicity of writing. The contact term does not contribute to the stress for $\bm{r}_0\neq \bm{r}$, but it is required for mathematical consistency as we shall see below. When substituted in Eq.~(\ref{eq:G}) $\varepsilon\kappa^2 \mathrm{G}_C$ immediately gives the $\bm{Y}_{lm}\otimes\bm{Y}^*_{lm}$ and $\bm{\Psi}_{lm}\otimes\bm{\Psi}^*_{lm}$ terms for $\delta(\bm{r}-\bm{r}_0)\mathbb{1}$. Furthermore, we obtain from the curl identities [Eqs.~(3.12) of Barrera {\it et al.} \cite{Barrera}]:
\begin{eqnarray}
 \nabla\times \mathrm{G}_C & = & \frac{1}{\kappa^2\varepsilon(r_0)}\sum_{lm} \bm{\Phi}_{lm}\otimes \bigg(-\frac{\delta(r-r_0)}{r^2 r_0} \bm{Y}_{lm}^*
 \nonumber\\
 && \quad\quad\quad\quad\,  + \frac{1}{r} \partial_r \frac{\delta(r-r_0)}{r_0} \frac{\bm{\Psi}_{lm}^*}{l(l+1)}\bigg)
 \end{eqnarray}
and writing $r^{-1}\partial_r r_0^{-1}\delta(r-r_0)$ as $-r_0^{-1}\partial_{r_0}\delta(r-r_0)$ and using the curl identities once more, but this time for $\bm{r}_0$:
\begin{equation}
\nabla\times\mathrm{G}_C = \frac{\delta(r-r_0)}{\kappa^2 r r_0\,\varepsilon(r_0)}\sum_{lm} \bm{\Phi}_{lm}\otimes \bm{\Phi}_{lm}^* \times\overleftarrow{\nabla}_0 \,.
\end{equation}
Requiring  Eq.~(\ref{eq:g}) for $g_M^{lm}$ (with $\nu=\varepsilon$) we obtain along similar lines as in the case of the electric polarization:
\begin{equation}
\nabla\times\frac{1}{\mu}\nabla\times (\mathrm{G}_M + \mathrm{G}_C ) +  \varepsilon \kappa^2 \mathrm{G}_M = 0 \,.
\end{equation}
This proves that, in Eq.~(\ref{eq:G})  for the Green tensor, $\mathrm{G}_M$ plus $\mathrm{G}_C$ generates the remaining terms in Eq.~(\ref{eq:delta}) for the right--hand side. 

Equation (\ref{eq:g}) thus defines the scalar Green functions for both polarizations. As Eq.~(\ref{eq:g}) does not depend on $m$ the Green function components $g^{lm}$ do not depend on $m$ either. The scalar Green functions are given in terms of the components $g^{lm}$ with respect to the standard spherical harmonics. In position space they read as
\begin{equation}
g(\bm{r},\bm{r_0})= \sum_{lm}g^{lm}(r,r_0)\,Y_{lm}(\theta,\phi)\,Y^*_{lm}(\theta_0,\phi_0) 
\label{eq:sphharmonics}
\end{equation}
and the scalar wave equation (\ref{eq:g}) takes the form of Eq.~(\ref{wave}) with definition (\ref{g}).

\subsection{Electric and magnetic stresses}

The fluctuation--dissipation theorem \cite{Scheel} relates the vacuum stress to the matrix Green function. In particular, defining 
\begin{eqnarray}
\tau_E &=& \frac{1}{2}\langle 0| \bm{D}(\bm{r})\otimes \bm{E}(\bm{r}_0) +  \bm{D}(\bm{r}_0)\otimes \bm{E}(\bm{r}) |0\rangle \,, \nonumber\\
\tau_M &=& \frac{1}{2}\langle 0| \bm{B}(\bm{r})\otimes \bm{H}(\bm{r}_0) +  \bm{B}(\bm{r}_0)\otimes \bm{H}(\bm{r}) |0\rangle
\end{eqnarray}
we arrive at the expressions \cite{Scheel,LeoBook}:
\begin{eqnarray}
\tau_E&=&-\frac{\hbar c}{\pi}\int_0^\infty \kappa^2\varepsilon\, (\mathrm{G}_E+\mathrm{G}_M)\,d\kappa \,,\nonumber\\
\tau_M&=&\frac{\hbar c}{\pi}\int_0^\infty\frac{1}{\mu}\nabla\times(\mathrm{G}_E+\mathrm{G}_M)\times\overleftarrow{\nabla}_0\,d\kappa \,.
\label{eq:taus}
\end{eqnarray}
The stress tensors are then given by
\begin{equation}
\sigma_F = \tau_F - \frac{1}{2}\,\mathrm{tr}\,\tau_F\,\mathbb{1}
\label{eq:tau}
\end{equation}
in the limit $\bm{r}_0\rightarrow\bm{r}$. In order to derive manageable expressions for the stress--tensor contributions, we write them in the spectral representation (\ref{stress}) for the components $i$, fields $F$ and polarizations $P$ as
\begin{equation}
W^{i P}_{i F}= \sum_{lm} {\cal W}_{i F}^{i P} \, Y_{lm}(\theta,\phi)Y^*_{lm}(\theta_0,\phi_0) \,.
\label{eq:calW}
\end{equation}
The sum of the angular coefficients ${\cal W}_{i F}^{i P}$ over $P$ corresponds to the Fourier component ${\cal W}$ in the planar case \cite{Grin1}. To illustrate our method of determining the ${\cal W}_{i F}^{i P}$ consider the simplest case: ${\cal W}_{i E}^{i E}$. We get for the corresponding part of the correlation function $\tau_E$:
\begin{eqnarray}
\tau_E^{(E)} &=& -\frac{\hbar c}{\pi}\int_0^\infty \kappa^2\varepsilon\, \mathrm{G}_E \,d\kappa \nonumber\\
&=& \frac{\hbar c}{\pi} \sum_{lm} \int_0^\infty \frac{\kappa^2\varepsilon g^{lm}_E}{l(l+1)}\,\bm{\Phi}_{lm}\otimes\bm{\Phi}^*_{lm}
\end{eqnarray}
from Eq.~(\ref{eq:GEM}) and assuming the limit $\theta_0\rightarrow\theta$ and $\phi_0\rightarrow\phi$. Then we use the fact that $g^{lm}$ is independent of $m$ and apply the last one of the sum identities (\ref{sumidentities}). From this and Eq.~(\ref{eq:tau}) follows
\begin{equation}
{\cal W}_{rE}^{rE} = \kappa^2\varepsilon\, g_{E}^{lm} \,.
\label{eq:WE}
\end{equation}
For calculating ${\cal W}_{i M}^{i E}$ we define the corresponding $\tau_M^{(E)}$ according to Eq.~(\ref{eq:taus}), use Eq.~(\ref{eq:GEM}), apply the curl identities of the vector spherical harmonics [Eqs.~(3.12) of Barrera {\it et al.} \cite{Barrera}] and finally use the sum identities (\ref{sumidentities}). We obtain from Eq.~(\ref{eq:tau}):
\begin{eqnarray}
{\cal W}_{rM}^{rE}&=&-\frac{\partial_{r_0}r_0\,\partial_rr \,g_{E}^{lm}}{\mu rr_0} - {\cal W}_{\theta M}^{\theta E} \,,\nonumber\\
{\cal W}_{\theta M}^{\theta E}&=&{\cal W}_{\phi M}^{\phi E}=-\frac{l(l+1)}{\mu rr_0}\,g^{lm}_{E} \,.
\label{eq:WM}
\end{eqnarray}
Equations (\ref{eq:WE}) and (\ref{eq:WM}) give the expressions in Eq.~(\ref{spectralstresses}) when represented in position space according to Eqs.~(\ref{eq:sphharmonics}) and (\ref{eq:calW}). In order to determine the components for the magnetic polarization we may proceed similarly, but applying also the wave equation (\ref{eq:g}) to reduce the number of derivatives \cite{MagRemark}. Alternatively, we may just take advantage of the electromagnetic duality and interchange the fields and the polarizations $E\leftrightarrow M$ and $\mu\leftrightarrow\varepsilon$.

\section{Geometrical optics}

In this appendix we provide some more details on the geometrical--optics theory for the renormalizing scalar Green functions $D_0$ and $D_1$. We closely follow the planar case \cite{Grin1} where the general expressions have been derived, and state the results for the spherical case. 

\subsection{Quadratic expansion}

As renormalization is local, the renormalizing Green function $D$ should depend on the local $\varepsilon$ and $\mu$ and their derivatives. We postulate that these are maximally second derivatives. This means that we expand the dielectric functions up to second order around the point of emission $\bm{r}_0$:
\begin{equation}
n = n_0 + n_0' (r-r_0) + \frac{n_0''}{2}(r-r_0)^2
\label{expansion}
\end{equation}
with $n_0=n(r_0)$, and similarly for $\varepsilon$ and $\mu$. The renormalizing Green function thus perceives the medium as if the dielectric functions depend on the point of emission at $r_0$, which potentially breaks reciprocity.

\subsection{Optical length}

The central quantity of geometrical optics is the optical length $s$ that satisfies the eikonal equation:
\begin{equation}
(\nabla s)^2 = n^2 \,.
\label{eikonal}
\end{equation}
Spherical symmetry implies that $s$ depends on $r$, $r_0$ and the angle $\gamma$ between the origin, $\bm{r}$ and $\bm{r}_0$. The distance between $\bm{r}$ and $\bm{r}_0$ is given by:
\begin{equation}
\rho=\sqrt{r^2+r_0^2 -2rr_0\cos\gamma} \,.
\end{equation}
One verifies that the solution of Eq.~(\ref{eikonal}) is: 
\begin{eqnarray}
s &=& \rho \left(n_0 + \frac{n_0'}{2} (r-r_0) +\frac{n_0''}{6}(r-r_0)^2 \right. \nonumber\\
&& \left. \quad\quad -\frac{n_0'r_0\, (n_0'r_0+2n_0)}{24n_0} \,\sin^2\gamma\right)
\label{sex}
\end{eqnarray}
up to quadratic order in $r-r_0$ and $\gamma$. One also verifies the reciprocity of $s$: if we replace $r\leftrightarrow r_0$ and $n\leftrightarrow n_0$ in Eq.~(\ref{sex}) and use expansion (\ref{expansion}) we obtain the same $s$, apart from third--order corrections. So despite the fact that $n$ depends on $r_0$, the optical length $s$ is reciprocal up to quadratic order.

\subsection{Amplitude}

The amplitude ${\cal A}_0$ of the wave in geometrical optics satisfies the continuity equation \cite{Grin1}:
\begin{equation}
\nabla \cdot \left(\nu {\cal A}_0^2 \nabla s \right) = 0
\label{cont}
\end{equation}
with $\nu$ defined in Eq.~(\ref{nu}). One verifies that up to quadratic order the solution of Eq.~(\ref{cont}) is the same as in the planar case \cite{Grin1}:
\begin{equation} 
{\cal A}_0 = -\frac{1}{4\pi\sqrt{\nu_0\nu}}\left(\frac{1}{\rho} + \frac{n_0^2 R_0}{48}\rho\right)
\label{ampex}
\end{equation}
where $R_0$ denotes the 3D curvature scalar \cite{LeoPhil}:
\begin{equation} 
R_0 = -\frac{4\nabla^2 n_0}{n_0^3} + \frac{2n_0'^2}{n_0^4} \,.
\label{curvature}
\end{equation}
The only difference is that in the planar case \cite{Grin1} $\nabla^2=\partial_z^2$ whereas in the spherical case we have $\nabla^2=\partial_r^2+(2/r)\partial_r$. One verifies again that ${\cal A}_0$ is reciprocal up to quadratic order.

\subsection{First scattering}

As $s$ and ${\cal A}_0$ are reciprocal, the outgoing wave $D_0$ defined in Eq.~(\ref{D0}) is reciprocal, too. The contribution $D_1$ of Eq.~(\ref{D1}) depends on $s$ and ${\cal A}_0$ as well, but also on the amplitude $\beta_1$. This quantity describes the first scattering in inhomogeneities around the point of emission \cite{Grin1}. The scattering amplitude is given by \cite{Grin1}:
\begin{equation} 
\beta_1 = \frac{\nabla\cdot\nu\nabla{\cal A}_0}{2n^2\nu{\cal A}_0} - \frac{R_r^r}{4} 
\label{betaex}
\end{equation}
with $R_r^r$ defined in Eq.~(\ref{ricci}). For $\nu=n$ when the medium establishes an exact geometry, $R_r^r$ is the radial component of the 3D Ricci tensor \cite{LeoPhil}. Evaluating Eq.~(\ref{betaex}) we obtain Eq.~(\ref{beta}) for $\beta_1$. One verifies that $D=D_0+D_1$ with the expressions given satisfies the wave equation (\ref{wave}) around $\bm{r}_0$ within $\mathrm{O}(\kappa^{-1})$. The scattering amplitude $\beta_1$ depends entirely on $r_0$ and therefore clearly breaks reciprocity, which causes the van der Waals anomaly.

\section{Asymptotics}

In order to see whether the renormalizing Green functions remove the infinity of the vacuum stress, we need to identify and characterize this divergency in the first place.  The stress is determined in Eq.~(\ref{stress}) by an integration over spectral stresses, and those are given in Eq.~(\ref{spectralstresses}) by maximally second derivatives of the scalar Green functions $g_P$. We thus need to analyse the asymptotics of the $g_P(\bm{r},\bm{r}_0)$ for large $\kappa$ and $\bm{r}_0\sim\bm{r}$. The Green functions $g_P$ we represent in Eq.~(\ref{eq:sphharmonics}) by the Green coefficients $g^{lm}$ dropping from now on the polarization label $P$. The required asymptotics of $g$ is then given by the asymptotics of the $g^{lm}$ for large $l$ and large $\kappa$. This is the asymptotics we determine in this appendix. 

\subsection{Langer transformation}

Similar to the planar case \cite{Grin1} we apply a WKB technique. However, for the spherical case, it is wise to adopt the Langer transformation \cite{Langer}
\begin{equation} 
r = e^x \,,\quad g^{lm} = e^{-x/2-x_0/2} \,u(x)
\label{eq:langer}
\end{equation}
dropping in $u$ the $lm$--indices for simplicity. We have put the constant $x_0/2$ in the exponent for keeping the reciprocity of $g^{lm}$ in $u$. Equation (\ref{eq:g}) appears then in the form:
\begin{equation} 
\partial_x \frac{1}{\nu}\, \partial_x u -\frac{\nu' u}{2\nu^2} - \frac{p^2 +e^{2x} \kappa^2 n^2}{\nu} u = \delta(x-x_0) 
\label{eq:lang}
\end{equation}
where the primes denote differentiations with respect to $x$ (such that $\nu'=r\partial_r \nu$) and the parameter $p$ abbreviates
\begin{equation} 
p=l+\frac{1}{2} 
\label{eq:index}
\end{equation}
that should not be confused with the pressure. In the following we determine the solution of Eq.~(\ref{eq:lang}) for large $\kappa$ and $p$. 

\subsection{Wronskian representation}

First, we write down an expression for the exact solution in terms of the homogeneous solutions $h_\pm$ of Eq.~(\ref{eq:lang}):
\begin{equation} 
u(x,x_0)=\frac{1}{W}
	\begin{cases}
		h_+(x)h_-(x_0) :& x>x_0\\
		h_+(x_0)h_-(x) :& x<x_0
	\end{cases}
\label{eq:uexact}
\end{equation}
where $W$ denotes the Wronskian:
\begin{equation}
W=\frac{h'_+(x)\,h_-(x) - h'_-(x)\,h_+(x)}{\nu} \,.
\label{eq:wronskian}
\end{equation}
The Wronskian is constant, as a consequence of Eq.~(\ref{eq:lang}). Since the $h_\pm$ solve Eq.~(\ref{eq:lang}) for $r\neq r_0$, $u$ does this as well. One verifies that the jump at $x=x_0$ specified in Eq.~(\ref{eq:uexact}) generates the delta function on the right hand side of Eq.~(\ref{eq:lang}), which proves that $u$ solves Eq.~(\ref{eq:lang}). The $h_\pm$ need to be chosen such that they decay at $\pm\infty$, respectively, which determines $u$ and hence the Green coefficient $g^{lm}$ uniquely. 

\subsection{WKB asymptotics}

Now we use the standard WKB technique for the homogeneous solutions $h_\pm$ in the limit $\kappa,p \rightarrow\infty$. For convenience, we replace $\kappa$ and $p$ by $\kappa/q$ and $p/q$ in terms of the formal parameter $q$ we regard as small. We write 
\begin{equation}
\label{eq:WKBansatz}
h_\pm=\exp\left(-\frac{1}{q}\sum_{m=0}^\infty q^m s_m\right) 
\end{equation}
and solve, order by order, the homogeneous equation
\begin{equation} 
\partial_x \frac{1}{\nu}\, \partial_x h -\frac{\nu' h}{2\nu^2} - \frac{p^2 +e^{2x} \kappa^2 n^2}{q^2\nu} u = 0 \,.
\label{eq:homo}
\end{equation}
Inserting our ansatz (\ref{eq:WKBansatz}) into Eq.~(\ref{eq:homo}), we get the relation:
\begin{gather}
\frac{1}{q^2}\sum_{m=0}^\infty q^m \sum_{k=0}^m \frac{s'_k s'_{m-k}}{\nu} 
+ \frac{1}{q}\sum_{m=0}^\infty q^m \left(-\frac{1}{\nu}s''_m+\frac{\nu'}{\nu^2}s'_m\right) \nonumber\\
= \frac{p^2 +e^{2x}\kappa^2n^2}{q^2\nu}+\frac{\nu'}{2\nu^2} \,.
\label{eq:rec0}
\end{gather}
In lowest order ($q^{-2}$) we have
\begin{equation}
s_{0}' = \pm  \sqrt{ \kappa^2n^2 e^{2x} + p^2}
\label{eq:s0}
\end{equation}
where the $\pm$ corresponds to the two choices $h_\pm$ that should and do vanish for $x\rightarrow\pm\infty$ according to Eq.~(\ref{eq:WKBansatz}). In the second--lowest order ($q^{-1}$) we have
\begin{equation}
s_{1}' = \frac{1}{2s_0'}\left(s''_{0}-\frac{\nu'}{\nu}s'_{0}\right) ,
\label{eq:s1}
\end{equation}
whereas in the next order ($q^0$) we get
\begin{equation}
s'_2=\frac{1}{2s'_0}\left(s''_{1}-\frac{\nu'}{\nu}s'_{1}-{s_1'}^2 - \frac{\nu'}{2\nu}\right).
\label{eq:s2}
\end{equation}
For all other orders, Eq.~(\ref{eq:rec0}) reduces to
\begin{equation}
s'_m=\frac{1}{2s'_0}\left(s''_{m-1}-\frac{\nu'}{\nu}s'_{m-1}-\sum_{k=1}^{m-1}s'_k s'_{m-k}\right)
\label{eq:rec}
\end{equation}
that includes Eq.~(\ref{eq:s1}) as a special case. Note that Eq.~(\ref{eq:rec}) is exactly the same recurrence relation as in the planar case \cite{Grin1}. Note also that the $\pm$ in Eq.~(\ref{eq:s0}) does not change the sign of $s_1'$ in Eq.~(\ref{eq:s1}). The $\pm$ sign carries then over from $s_0'$ to $s_2'$ in Eq.~(\ref{eq:s2}). In fact, all even orders carry the $\pm$ sign, while all odd orders are unchanged, as one sees from Eq.~(\ref{eq:rec}). In terms of our ansatz (\ref{eq:WKBansatz}) we thus have
\begin{equation}
h_\pm = e^{\mp s_{\cal E}} e^{- s_{\cal O}}
\label{eq:hpm}
\end{equation}
where the $s_{\cal E}$ and $s_{\cal O}$ collect the even and orders in $s$, respectively:
\begin{equation}
s_{\cal E} \equiv \frac{1}{q}\sum_{m=0}^\infty q^{2m}s_{2m} \,,\quad
s_{\cal O} \equiv \frac{1}{q}\sum_{m=0}^\infty q^{2m+1}s_{2m+1}.
\end{equation}
Given our result (\ref{eq:hpm}) for $h_\pm$, we write down the Wronskian of Eq.~(\ref{eq:wronskian}):
\begin{equation}
W=-\frac{2}{\nu}\,s_{\cal E}'\,e^{-2s_{\cal O}}.
\end{equation}
As $W=\mathrm{const}$, we may put $W=\sqrt{W(x)W(x_0)}$ and have
\begin{equation}
W=-2\sqrt{\frac{s_{\cal E}'(x)s_{\cal E}'(x_0)}{\nu(x)\nu(x_0)}} \, e^{-s_{\cal O}(x)-s_{\cal O}(x_0)} \,,
\end{equation}
which gives the compact expression
\begin{equation}
u=-\frac{1}{2}\sqrt{\frac{\nu(e^x)\nu(e^{x_0})}{s_{\cal E}'(x)s_{\cal E}'(x_0)}}\exp\left(-\abs{\int_{x_0}^x s_{\cal E}' \,dx}\right)
\label{eq:uasy}
\end{equation}
that only depends on the even orders in the asymptotic expansion (\ref{eq:WKBansatz}). Formula (\ref{eq:uasy}) is explicitly reciprocal, and so is the Green function given by the Langer transformation, Eq.~(\ref{eq:langer}). Note that Eq.~(\ref{eq:uasy}) is identical with the expression for the Fourier--transformed Green functions in the planar case \cite{Grin1}. However, the starting point of the recurrence and the expression for $s_2$, Eqs.~(\ref{eq:s0}) and (\ref{eq:s2}), are different. 

\subsection{Divergency of the stress}

Now we are going to characterize the divergency of the vacuum stress using as cutoff $\Lambda=1/q$ and the expressions derived for the scalar Green function. But first we note that due to dispersion \cite{LL8} the refractive index and impedance have the following asymptotic form in the high--frequency limit:
\begin{equation}
n(r;\kappa)\sim1+\frac{n_\infty(r)}{\kappa^2}\,,\quad Z(r;\kappa)\sim 1 + \frac{Z_\infty(r)}{\kappa^2} 
\label{eq:dispersion}
\end{equation}
where the impedance is defined as $Z=\sqrt{\mu/\varepsilon}$. 

In the following we express our results in terms of the spectral stresses defined in Eq.~(\ref{eq:calW}). We take dispersion into account according to Eq.~(\ref{eq:dispersion}) and calculate the bare stresses given by Eqs.~(\ref{eq:WE}) and (\ref{eq:WM}) using our WKB technique. We obtain for ${\cal W}^r_r$ for each order of divergency:
\begin{eqnarray}
\Lambda^4 & : & -\frac{4 p w}{r^3} \,, \nonumber\\
\Lambda^2 & : & \frac{p \left(-8 r^2 w^4 n_\infty+p^4+w^4\right)}{2
r^3 w^5} \,, \nonumber\\
\ln \Lambda & : & \frac{p}{32 r^3 w^{11}\left(w^2-p^2\right)} \Bigl(32
p^4 r^3 w^6 n_\infty' \nonumber\\
& & - 16 r^2 w^4 n_\infty \left(w^2-p^2\right) \left(5 p^4+w^4\right)\nonumber\\
& & - \left(w^2-p^2\right)^2  \bigl(105 p^6-63 p^4 w^2+7 p^2
w^4-w^6\bigr) \nonumber\\
& & - \,64 p^2 r^4 w^8 n_\infty^2 \Bigr)
\label{eq:rr}
\end{eqnarray}
with the abbreviation
\begin{equation}
w=\sqrt{p^2+\kappa^2r^2} \,.
\end{equation}
We obtain for ${\cal W}^\theta_\theta$:
\begin{eqnarray}
\Lambda^4 & : & \frac{2 p^3}{r^3 w} \,, \nonumber\\
\Lambda^2 & : & \frac{-8 p^3 r^2 w^4 n_\infty+5 p^7-6 p^5 w^2+p^3
w^4-2 p w^6}{4 r^3 w^7} \,, \nonumber\\
\ln \Lambda & : & \frac{p}{64 r^3 w^{13} \left(w^2-p^2\right)} \Bigl(
-(w^2-p^2)^2 (1155 p^8 \nonumber\\
& & \quad - 1617 p^6 w^2+553 p^4 w^4-47 p^2 w^6+4 w^8) \nonumber\\
& & + 16 r^2 w^4 (2 p^4 r w^2 (r w^2 n_\infty''-5 (w^2-p^2) n') \nonumber\\
& & \quad + n (35 p^8 -65 p^6 w^2+33 p^4 w^4-5 p^2 w^6+2 w^8) \nonumber\\
& & \quad + 4 p^2 r^2 (2w^2-3 p^2) w^4 n_\infty^2 )\Bigr) \,.
\label{eq:thth}
\end{eqnarray}
These are the divergencies of the vacuum stress. In Appendix D we prove that they are removed by renormalization. 

\section{Convergence}

In this appendix we prove that the renormalization converges. By this we mean that the infinite vacuum stress calculated from the renormalizer, $D_0$ +$D_1$, compensates for the infinite vacuum stress calculated using the asymptotics of the Green function (Appendix C). What is left is finite. 

\subsection{Spherical harmonics decomposition}

In order to compare the vacuum stress of the renormalizer with the bare stress given by the asymptotics of the Green function, we need to represent $D_0$ and $D_1$ in terms of spherical harmonics.  Since we are only interested in the behavior up to quadratic order, we approximate and decompose $D_0$ and $D_1$ according to Eqs.~(\ref{D0}), (\ref{D1}), (\ref{expansion}), (\ref{sex}) and (\ref{ampex}):
\begin{eqnarray}
 D_0 & = &{\cal A}_0\,e^{-\kappa s}\nonumber\\
&\sim & - \frac{1}{4\pi\rho\sqrt{\nu\nu_0}}\left[1+\frac{n_0^2R_0}{48}\rho^2\right](1+\kappa\rho\alpha_0\gamma^2)\,e^{-\kappa \rho\chi}\nonumber\\
&=&\sum_{m=1}^4D_0^m \quad\mbox{where} \label{eq:dseries} \\
\alpha_0 &=& \frac{n_0'r_0\, (n_0'r_0+2n_0)}{24n_0} \,,\\
\chi &=& n_0 + \frac{n_0'}{2} (r-r_0) +\frac{n_0''}{6}(r-r_0)^2 \,, \\
D^1_0 & = & -\frac{1}{4\pi\rho\sqrt{\nu\nu_0}} \, e^{-\kappa  \rho\chi}\,,\\
D^2_0 & = & -\frac{1}{4\pi\sqrt{\nu\nu_0}}\frac{n_0^2R_0}{48}\rho \, e^{-\kappa  \rho\chi}\,,\\
D^3_0 & = & -\frac{1}{4\pi\sqrt{\nu\nu_0}}\kappa\alpha_0\gamma^2 \, e^{-\kappa  \rho\chi}\,,\\
D^4_0 & = &-\frac{1}{4\pi\sqrt{\nu\nu_0}}\frac{n_0^2R_0}{48}\rho^2\kappa\alpha_0\gamma^2 \, e^{-\kappa  \rho\chi}\,,\\
D_1&= & {\cal A}_0\frac{\beta_1 s}{\kappa}e^{-\kappa s}\sim - \frac{1}{4\pi\sqrt{\nu\nu_0}}\frac{\beta_1}{\kappa} \, e^{-\kappa  \rho\chi} \,. 
\label{eq:dseries1}
\end{eqnarray}
These terms generate all divergencies in the vacuum stress, with further terms being of order $\mathrm{O}(\Lambda^{-2})$ or below in their contribution to the stress.

Equations (\ref{eq:dseries}-\ref{eq:dseries1}) reduce our problem to figuring out how to decompose the functions $f$ and  $\gamma^2f_\alpha$ into spherical harmonics with
\begin{equation}
f_\alpha(\bm{r};\bm{r}_0)=\frac{\rho^\alpha}{4\pi}e^{-k\rho}
\label{eq:fdef}
\end{equation}
where $k$ abbreviates $\kappa\chi$ here. The $f_\alpha$ depend via $\rho$ on the angles $\theta$ and $\phi$ of the spherical coordinates, whereas $k$ does not depend on them. The spherical harmonics $f_{-1,lm}$ of $f_{-1}$ are given by the well--known expression \cite{Erdelyi}:
\begin{equation}
	f_{-1,lm}(r,r_0;k)=
	\frac{1}{\sqrt{rr_0}}\begin{cases}
		K_p (k r_0)I_p (k r) &r<r_0\\
		I_p (k r_0)  K_p (k r) &r>r_0
	\end{cases}
\label{eq:fminus1}
\end{equation}
where $I_p$ and $K_p$ are the modified Bessel functions \cite{Erdelyi} with indices $p=l+\frac{1}{2}$. From definition (\ref{eq:fdef}) follows then
\begin{equation}
f_{\alpha,lm}=(-k)^{-\alpha-1}\partial_k^{\alpha+1}f_{-1,lm} \,.
\end{equation}
Given the decomposition of $f_\alpha$ one can also determine the spherical harmonics for $\gamma^2f_\alpha$ and $\gamma\sim 0$. For this we take $\gamma^2\sim 2(1-\cos\gamma)$ and rotate our coordinate system such that $\gamma=\theta$. From the recurrence relations of the Legendre polynomials\cite{Erdelyi} in the spherical harmonics \cite{Erdelyi} follows then:
\begin{equation}
\label{eq:cgf}
(\cos\gamma f)_{lm}=\frac{(l+1) f_{l+1,m}-l f_{l-1,m}}{2l+1} \,.
\end{equation}
Now we need to determine the asymptotics of these expressions for large $\kappa$ and $p$ with $p=l+\frac{1}{2}$.

\subsection{Asymptotics}

As the vacuum stress is given by a linear differential operator on the Green function, its asymptotic behavior for $\bm{r}_0\rightarrow\bm{r}$ is determined by this linear differential operator applied on functions of type $IK$ and their derivatives. We take the choice $r>r_0$ (or $r<r_0$) in Eq.~(\ref{eq:fminus1}). Using the Bessel identities \cite{Erdelyi}
\begin{eqnarray}
  \frac{2\alpha}{x}I_\alpha(x)&=&I_{\alpha-1}(x)-I_{\alpha+1}(x) \,,\nonumber\\
  2I_\alpha'(x)&=&I_{\alpha-1}(x)+I_{\alpha+1}(x)\,,\nonumber\\
  -\frac{2\alpha}{x}K_\alpha(x)&=&K_{\alpha-1}(x)-K_{\alpha+1}(x)\,,\nonumber\\
  -2K_\alpha'(x)&=&K_{\alpha-1}(x)+K_{\alpha+1}(x)
\end{eqnarray}
we can always reduce the stress to the form:
\begin{eqnarray}
\label{eq:Rs}
\sigma &\sim& R_1 I_{p}(\kappa \chi r_0)K_{p}(\kappa \chi r) \nonumber\\
&&  + R_2I_{p+1}(\kappa \chi r_0)K_{p}(\kappa \chi r)  \nonumber\\
&& + R_3I_{p}(\kappa \chi r_0)K_{p+1}(\kappa \chi r) \nonumber\\
&& +R_4I_{p+1}(\kappa \chi r_0)K_{p+1}(\kappa \chi r)
\end{eqnarray}
where the $R_i$ are some rational functions of $r,r_0,\kappa$ and $p$. We only need the asymptotics of the products $IK$ for large index and large argument.

The required uniform asymptotic behaviour of the Bessel functions is well understood. For large $\alpha$ and finite $z$ we have \cite{Olver}:
\begin{eqnarray}
I_\alpha(\alpha z) &\sim& \frac{e^{\alpha\xi}}{(2\pi\alpha)^{1/2}(1+z^2)^{1/4}}\sum_{s=0}^\infty\frac{U_s(w)}{\alpha^s}\,, \nonumber\\
K_\alpha(\alpha z) &\sim& \left(\frac{\pi}{2\alpha}\right)^{1/2}\frac{e^{-\alpha\xi}}{(1+z^2)^{1/4}}\sum_{s=0}^\infty(-1)^s\frac{U_s(w)}{\alpha^s} \,, \nonumber\\
\xi &=& (1+z^2)^{1/2}+\ln\frac{z}{1+(1+z^2)^{1/2}} \,,\nonumber\\
w &=& (1+z^2)^{-1/2}
\end{eqnarray}
in terms of the polynomials $U_s$ defined by the relations
\begin{eqnarray}
U_{s+1}(x)&=&\frac{1}{2} x^2(1-x^2)U_s'(x)+\frac{1}{8}\int_0^x(1-5t^2)U_s(t) \,d t \,,\nonumber\\
U_0(x)&=&1 \,.
\end{eqnarray}
This gives the asymptotics of the $IK$ products:
\begin{align}
&I_{\alpha_1}(\alpha_1 z_1)K_{\alpha_2}(\alpha_2 z_2) \sim \frac{e^{\alpha_1\xi_1-\alpha_2\xi_2}}{2(\alpha_1\alpha_2)^{1/2} \,(1+z_1^2)^{1/4} \, (1+z_2^2)^{1/4}}
\nonumber\\
& \qquad \times \sum_{\substack{0\le s_1< \infty\\0\le s_2\le s_1}} (-1)^{s_2}\frac{U_{s_1-s_2}(w_1)}{\alpha_1^{s_1-s_2}}\frac{U_{s_2}(w_2)}{\alpha_2^{s_2}} \,.
\label{eq:IK}
\end{align}
We use this equation to get the asymptotic forms of all the needed $IK$ combinations, by taking  $\alpha_1,\alpha_2=p,p+1$ as necessary, and $z_1=(\kappa\chi r)/p $ for the $p$ case and $z_1=(\kappa \chi r_0)/(p+1)$ for the $(p+1)$ case. From here we expand around \(\Lambda \to \infty\), take \(r\to r_0\), and obtain the divergency. The calculations are best done with computer algebra; in the following we state the main result. 

\subsection{Result}

We find that the $D_0$ contribution to the stress almost takes care of all the infinities described in Eqs.~(\ref{eq:rr}) and (\ref{eq:thth}) with only one term left:
\begin{equation}
 \left. \mathcal{W}^{\theta}_{\theta}-\mathcal{W}^{\theta}_{\theta}\right|_{D_0} =  \frac{p^3}{w^3}\,\frac{r n''_\infty(r)-n'_\infty(r)}{3 \kappa^2 r^2} \,.
 \label{eq:gap}
\end{equation}
If we think of the $p$--summation as an integration and write $p=w\cos\vartheta$ and $\kappa r=w\sin\vartheta$ such that we get a two--dimensional integral in polar coordinates with (large) radius $w$, the result (\ref{eq:gap}) leads to a logarithmic divergence in $\Lambda$. Using Eq.~(\ref{eq:dispersion}) we may also express the result as:
\begin{equation}
 \left. \mathcal{W}^{\theta}_{\theta}-\mathcal{W}^{\theta}_{\theta}\right|_{D_0} \sim  \frac{p^3}{w^3}\, \frac{2(\beta_E+\beta_M)}{r} 
 \label{eq:gap2}
\end{equation}
for large $\kappa$ in terms of the scattering amplitudes $\beta_E$ and $\beta_M$ defined in Eq.~(\ref{beta}). These scattering amplitudes characterize the $D_1$ waves of Eq.~(\ref{D1}). In fact, we find that the left--over term $ \left. \mathcal{W}^{\theta}_{\theta}-\mathcal{W}^{\theta}_{\theta}\right|_{D_0}$ is entirely accounted for by the contribution of $D_1$ to the stress. So, if we include $D_1$ in the renormalization procedure, the stress becomes finite.

\section{Vector spherical harmonics}

In this appendix we state the definitions, main relations and sum formulas for the vector spherical harmonics we use in this paper. 

\subsection{Definition}

According to Barrera {\it et al}. \cite{Barrera} the vector spherical harmonics are defined as:
\begin{eqnarray}
\bm{Y}_{lm}(\theta,\phi)&=&\vu{r} Y_{lm}(\theta,\phi)\,, \nonumber\\
\bm{\Phi}_{lm}(\theta,\phi)&=&\bm{r}\times\nabla Y_{lm}(\theta,\phi)\,, \nonumber\\
\bm{\Psi}_{lm}(\theta,\phi)&=& r\nabla Y_{lm}(\theta,\phi) 
\label{eq:defs}
\end{eqnarray}
where the $Y_{lm}$ are the usual spherical harmonics \cite{Erdelyi} and $\vu{r}=\bm{r}/r$. The vector spherical harmonics establish three orthogonal vectors in three--dimensional space for equal $lm$, and they are orthogonal functions on the unit sphere for different $lm$.

\subsection{Decomposition}

Any vector function $\bm{V}(\theta,\phi)$ on the unit sphere can be decomposed into vector spherical harmonics:
\begin{eqnarray}
\bm{V}(\theta,\phi)&=&\sum_{l=0}^\infty\sum_{m = -l}^{l}\Bigl(A_{lm}\bm{Y}_{lm}(\theta,\phi) + B_{lm}\bm{\Phi}_{lm}(\theta,\phi) \nonumber\\
&& \qquad\qquad +\, C_{lm}\bm{\Psi}_{lm}(\theta,\phi)\Bigr) 
\end{eqnarray}
with the coefficients given by \cite{Barrera}
\begin{eqnarray}
 A_{lm}&=& \int \bm{V}(\theta,\phi)\cdot \bm{Y^*}_{lm}(\theta,\phi) \, d\Omega \,, \nonumber\\
B_{lm}&=& \frac{1}{l(l+1)} \int \bm{V}(\theta,\phi)\cdot \bm{\Phi^*}_{lm}(\theta,\phi) \, d\Omega \,, \nonumber\\
C_{lm}&=& \frac{1}{l(l+1)} \int \bm{V}(\theta,\phi)\cdot \bm{\Psi^*}_{lm}(\theta,\phi) \, d\Omega \,.
\end{eqnarray}

\subsection{Sum identities}

In order to derive our results for the vacuum stress we make use of several sum identities of the vector spherical harmonics that are nigh impossible to find in the literature. Therefore we list them here. They originate from the familiar sum identity \cite{Erdelyi}
\begin{equation}
\frac{4\pi}{2l+1}\sum_{m=-l}^l Y_{lm}(\theta_1, \phi_1)\, Y^*_{lm}(\theta_2, \phi_2)=P_l(\cos\gamma)
\label{eq:sumid}
\end{equation}
where $\gamma$ denotes the angle between the two points $(\theta_1,\phi_1)$ and $(\theta_1,\phi_2)$ on the unit sphere, with
\begin{equation}
\cos\gamma = \cos\theta_1\cos\theta_2 + \sin\theta_1\sin\theta_2 \cos(\phi_1-\phi_2) \,.
\end{equation}
Let us explain our procedure by giving an example: the sum of $\bm{\Psi}_{lm}(\theta,\phi)\otimes\bm{\Psi}_{lm}(\theta,\phi)^*$ over $m$. From definition (\ref{eq:defs}) follows
\begin{align}
 & \bm{\Psi}_{lm}(\theta_1,\phi_1)\otimes\bm{\Psi}_{lm}(\theta_2,\phi_2)^*\nonumber\\
 &=\begin{bmatrix}
0&0&0\\
0&\partial_{\theta_1}\partial_{\theta_2}&\frac{1}{\sin\theta_2}\partial_{\theta_1}\partial_{\phi_2}\\
0&\frac{1}{\sin\theta_1}\partial_{\phi_1}\partial_{\theta_2}&\frac{1}{\sin\theta_1\sin\theta_2}\partial_{\phi_1}\partial_{\phi_2}\\
\end{bmatrix}
Y_{lm1} Y^*_{lm2}
\end{align}
with $Y_{lmi}\equiv Y_{lm}(\theta_i,\phi_i)$. Hence we obtain 
\begin{align}
  &\sum_{m=-l}^l \bm{\Psi}_{lm}(\theta_1,\phi_1)\otimes\bm{\Psi}_{lm}(\theta_2,\phi_2)^* \nonumber\\
  &=\begin{bmatrix}
0&0&0\\
0&\partial_{\theta_1}\partial_{\theta_2}&\frac{1}{\sin\theta_2}\partial_{\theta_1}\partial_{\phi_2}\\
0&\frac{1}{\sin\theta_1}\partial_{\phi_1}\partial_{\theta_2}&\frac{1}{\sin\theta_1\sin\theta_2}\partial_{\phi_1}\partial_{\phi_2}\\
\end{bmatrix}\frac{2l+1}{4\pi} \, P_l(\cos\gamma)
\end{align}
from Eq.~(\ref{eq:sumid}). We take the limit $\gamma\rightarrow 0$ and use for the $k$--th derivatives $P^{(k)}_l(1)$ the formula
\begin{equation}
P^{(k)}_l(1)=\frac{\prod_{q=-k+1}^k(l+q)}{\prod_{q=1}^k2q}
\end{equation}
that is easy to derive by induction. The result is given below, and also the results for similar sum identities we have derived with the same method:
\begin{align}
\sum_{m=-l}^l\bm{Y}_{lm}\otimes\bm{Y}_{lm}^*&=
\begin{bmatrix}
1&0&0\\
0&0&0\\
0&0&0\\
\end{bmatrix}
\frac{2l+1}{4\pi} 
\nonumber\\
\sum_{m=-l}^l\bm{Y}_{lm}\otimes\bm{\Psi}_{lm}^*&=0\,,\nonumber\\
\sum_{m=-l}^l\bm{Y}_{lm}\otimes\bm{\Phi}_{lm}^*&=0\,,\nonumber\\
\sum_{m=-l}^l\bm{\Psi}_{lm}\otimes\bm{Y}_{lm}^*&=0 \,,\nonumber\\
\sum_{m=-l}^l\bm{\Psi}_{lm}\otimes\bm{\Psi}_{lm}^*&=
\begin{bmatrix}
0&0&0\\
0&1&0\\
0&0&1\\
\end{bmatrix}
\frac{l(l+1)(2l+1)}{8\pi} \,,\nonumber\\
\sum_{m=-l}^l\bm{\Psi}_{lm}\otimes\bm{\Phi}_{lm}^*&=
\begin{bmatrix}
0&0&0\\
0&0&1\\
0&-1&0\\
\end{bmatrix}
\frac{l(l+1)(2l+1)}{8\pi} \,,\nonumber\\
\sum_{m=-l}^l\bm{\Phi}_{lm}\otimes\bm{Y}_{lm}^*&=0 \,,\nonumber\\
\sum_{m=-l}^l\bm{\Phi}_{lm}\otimes\bm{\Psi}_{lm}^*&=
\begin{bmatrix}
0&0&0\\
0&0&-1\\
0&1&0\\
\end{bmatrix}
\frac{l(l+1)(2l+1)}{8\pi}\,,\nonumber\\
\sum_{m=-l}^l\bm{\Phi}_{lm}\otimes\bm{\Phi}_{lm}^*&=
\begin{bmatrix}
0&0&0\\
0&1&0\\
0&0&1\\
\end{bmatrix}
\frac{l(l+1)(2l+1)}{8\pi} 
\label{sumidentities}
\end{align}
for $\theta_0=\theta$ and $\phi_0=\phi$.


\begin{thebibliography}{99}

\bibitem{DLP}
I. E. Dzyaloshinskii, E. M. Lifshitz, and L. P. Pitaevskii,
The general theory of van der Waals forces,
Adv. Phys. {\bf 10}, 165 (1961).

\bibitem{Buhmann}
S. Y. Buhmann,
{\it Dispersion Forces}
(Springer, Berlin, 2012),

\bibitem{Shahmoon}
E. Shahmoon,
{\it Van der Waals and Casimir--Polder dispersion forces},
in Ref.~\cite{Forces} below.

\bibitem{Forces}
W. M. R. Simpson and U. Leonhardt (eds.)
{\it Forces of the quantum vacuum}
(World Scientific, Singapore, 2015). 

\bibitem{Casimir}
H. B. G. Casimir, 
On the attraction between two perfectly conducting plates, 
Koninkl. Ned. Akad. Wetenschap. {\bf 51}, 793 (1948).

\bibitem{Rodriguez}
A. W. Rodriguez, F. Capasso, and S. G. Johnson, 
The Casimir effect in microstructured geometries,
Nat. Photon. {\bf 5}, 211 (2011).

\bibitem{CasimirPolder}
H. B. G. Casimir and D. Polder,
The Influence of Retardation on the London--van der Waals Forces,
Phys. Rev. {\bf  73}, 360 (1948).

\bibitem{Pita}
L. P. Pitaevskii,
Comment on ``Casimir force acting on magnetodielectric bodies embedded in media'',
Phys. Rev. A {\bf 73}, 047801 (2006).

\bibitem{Burger}
F. A. Burger, J. Fiedler, and S. Y. Buhmann,
Zero-point electromagnetic stress tensor for studying Casimir forces on colloidal particles in media,
Europhys. Lett. {\bf 121}, 24004 (2018). 

\bibitem{Levitation}
J. N. Munday, F. Capasso, and V. A. Parsegian,
Measured long--range repulsive Casimir--Lifshitz forces,
Nature {\bf 457}, 170 (2009).

\bibitem{CasimirEquilibrium}
R. Zhao, L. Li, S. Yang, W. Bao, Y. Xia, P. Ashby, Y. Wang, and X. Zhang,
Stable Casimir equilibria and quantum trapping,
Science {\bf 364}, 984 (2019).

\bibitem{Decca}
R. S. Decca,
{\it Measuring Casimir Phenomena},
in Ref.~\cite{Forces}.

\bibitem{SimpsonSurprise}
W. M. R. Simpson,
{\it Surprises in Theoretical Casimir Physics}
(Springer, Berlin, 2014).

\bibitem{Simpson}
W. M. R. Simpson, S. A. R. Horsley, and U. Leonhardt,
Divergence of Casimir stress in inhomogeneous media,
Phys. Rev. A {\bf 87}, 043806 (2013); {\it ibid.} {\bf 88}, 059901 (2013).

\bibitem{Grin1}
I. Griniasty and U. Leonhardt,
Casimir stress inside planar materials,
Phys. Rev. A {\bf 96}, 032123 (2017).

\bibitem{Grin2}
I. Griniasty and U. Leonhardt,
{\it Casimir stress in materials: Hard divergency at soft walls},
Phys. Rev. B {\bf 96}, 205418 (2017).

\bibitem{Milton80}
K. A. Milton,
Casimir Energy for a Spherical Cavity in a Dielectric: Toward a Model for Sonoluminescence?
Ann. Phys.  (New York) {\bf 127}, 49 (1980).

\bibitem{Brevik82}
I. Brevik and H. Kolbenstvedt,
The Casimir effect in a solid ball when $\varepsilon\mu= 1$,
Ann. Phys. (New York) {\bf 143}, 179 (1982).

\bibitem{Milton97}
K. A. Milton and Y. J. Ng,
Casimir energy for a spherical cavity in a dielectric: Applications to sonoluminescence,
Phys. Rev. E {\bf 55}, 4207 (1997).

\bibitem{Brevik98}
I. Brevik, V. V. Nesterenko and I. G. Pirozhenko, 
Direct mode summation for the Casimir energy of a solid ball,
J. Phys. A {\bf 31}, 8661 (1998).

\bibitem{Brevik99}
I. Brevik, V. N. Marachevsky, and K. A. Milton, 
Identity of the van der Waals Force and the Casimir Effect and the Irrelevance of These Phenomena to Sonoluminescence,
Phys. Rev. Lett. {\bf 82}, 3948 (1999).

\bibitem{Barton99}
G. Barton, 
Perturbative check on the Casimir energies of nondispersive dielectric spheres,
J. Phys. A {\bf 32}, 525 (1999).

\bibitem{Bordag99}
M. Bordag, K. Kirsten and D. Vassilevich, 
Ground state energy for a penetrable sphere and for a dielectric ball,
Phys. Rev. D {\bf 59}, 85011 (1999).

\bibitem{Avni}
Y. Avni and U. Leonhardt,
Casimir self--stress in a dielectric sphere,
Ann. Phys. (New York) {\bf 395}, 326 (2018).

\bibitem{Comment}
K. A. Milton, P. Parashar, I. Brevik, G. Kennedy,
Self--stress on a dielectric ball and Casimir--Polder forces,
Ann. Phys. (New York) {\bf 412}, 168008 (2020).

\bibitem{LL9}
L. D. Landau and E. M. Lifshitz,
{\it Statistical Physics, Part 2}
(Pergamon, Oxford, 1980).

\bibitem{Lifshitz}
E. M. Lifshitz, 
The Theory of Molecular Attractive Forces between Solids,
J. Exper. Theoret. Phys. USSR {\bf 29}, 94 (1954)
[English translation: Sov. Phys. JETP {\bf 2}, 73 (1956)].

\bibitem{Scheel}
S. Scheel,
{\it The Casimir stress in real materials},
in Ref.~\cite{Forces}. 

\bibitem{Wald}
R. M. Wald,
Trace anomaly of a conformally invariant quantum field in curved spacetime,
Phys. Rev. D {\bf 17}, 1477 (1978).

\bibitem{Weinberg}
S. Weinberg, 
The cosmological constant problem,
Rev. Mod. Phys. {\bf 61}, 1 (1989). 

\bibitem{HawkingInflation}
S. W. Hawking, T. Hertog, and H. S. Reall,
Trace anomaly driven inflation,
Phys. Rev. D {\bf 63}, 083504 (2001).

\bibitem{LL2}
L. D. Landau and E. M. Lifshitz,
{\it The Classical Theory of Fields}
(Butterworth-Heinemann, Amsterdam, 2003).

\bibitem{Dark}
L. Amendola and S. Tsujikawa,
{\it Dark Energy: Theory and Observations}
(Cambridge University Press, Cambridge, 2010).

\bibitem{Volovik}
G. Volovik, 
{\it The Universe in a Helium Droplet}
(Oxford University Press, Oxford, 2003).

\bibitem{Visser}
C. Barcelo, S. Liberati and M. Visser, 
Analogue Gravity, 
Living Rev. Relativity {\bf 8}, 12 (2005).

\bibitem{FF}
P. O. Fedichev and U. R. Fischer,
Gibbons--Hawking Effect in the Sonic de Sitter Space--Time of an Expanding Bose--Einstein--Condensed Gas,
Phys. Rev. Lett. {\bf 91}, 240407 (2003).

\bibitem{Kolo1}
E. B. Kolomeisky,
Natural analog to cosmology in basic condensed matter physics,
Phys. Rev. B {\bf 100}, 140301(R) (2019).

\bibitem{Kolo2}
E. B. Kolomeisky,
Normal modes of vibrations around Hubble flow in jellium,
Phys. Rev. B {\bf 101}, 174304 (2020).

\bibitem{Kolo3}
E. B. Kolomeisky,
Analog de Sitter space in a controlled Coulomb explosion,
Phys. Rev. A {\bf 103}, 043101 (2021).

\bibitem{Stein21}
J. Steinhauer, M. Abuzarli, T. Aladjidi, T. Bienaim\'{e}, C. Piekarski, W. Liu, E. Giacobino, A. Bramati, and Q. Glorieux,
Analogue cosmological particle creation in an ultracold quantum fluid of light,
arXiv:2102.08279.

\bibitem{Case}
U. Leonhardt, 
The case for a Casimir cosmology, 
Phil. Trans. R. Soc. A {\bf 378}, 20190229 (2020). 

\bibitem{Pethick}
C. Pethick and H. Smith,
{\it Bose--Einstein condensation in dilute gases}
(Cambridge University Press, Cambridge, 2008).

\bibitem{LL8}
L. D. Landau and E. M. Lifshitz,
{\it Electrodynamics of Continuous Media}
(Pergamon, Oxford, 1984).

\bibitem{CMBPlanck}
Planck Collaboration, 
Planck 2018 results. VI. Cosmological parameters,
Astron. \& Astrophys. {\bf 641}, A6 (2020).

\bibitem{Gordon}
W. Gordon,
Zur Lichtfortpflanzung nach der Relativit\"ats\-theorie,
Ann. Phys. (Leipzig) {\bf 72}, 421 (1923).

\bibitem{Plebanski}
J. Plebanski,
Electromagnetic Waves in Gravitational Fields,
Phys. Rev. {\bf 118}, 1396 (1960).

\bibitem{Schleich}
W. Schleich and M. O. Scully,
{\it General relativity and modern optics}, in
{\it New trends in atomic physics: Les Houches, session XXXVIII, 1982} by
G. Grynberg and R. Stora (eds.)
(Elsevier, Amsterdam, 1984).

\bibitem{LeoPhil}
U. Leonhardt and T. G. Philbin,
{\it Geometry and Light: the Science of Invisibility}, 
(Dover, Mineola, 2010).

\bibitem{Annals}
U. Leonhardt,
Lifshitz theory of the cosmological constant,
Ann. Phys. (New York)  {\bf 411}, 167973 (2019). 

\bibitem{Dror}
D. Berechya and U. Leonhardt,
Lifshitz cosmology: Hubble tension and quantum vacuum,
MNRAS {\bf 507}, 3473 (2021).

\bibitem{Verde}
L. Verde, T. Treu, and A. G. Riess,
Tensions between the early and late Universe,
Nat. Astron. {\bf 3}, 891 (2019).

\bibitem{Shell}
The famous, solved problem of the Casimir force on a perfectly conducting shell \cite{Boyer,MDS,Balian,Nesterenko} can be reduced to the problem of the Casimir force on the homogeneous sphere in a uniform background \cite{Avni}: the force on the shell is the force on a solid, perfectly conducting sphere plus the force on a hollow sphere in a uniform, perfectly conducting background.

\bibitem{Boyer}
T. H. Boyer,
Quantum electromagnetic zero-point energy of a conducting spherical shell and the Casimir model for a charged particle,
Phys. Rev. {\bf 174}, 1764 (1968).

\bibitem{MDS}
K. A. Milton, L. L. De Raad Jr, J. Schwinger,
Casimir self-stress on a perfectly conducting spherical shell,
Ann. Phys.  (New York) {\bf 115}, 388 (1978).

\bibitem{Balian}
R. Balian and B. Duplantier,
Electromagnetic waves near perfect conductors. I. Multiple scattering expansions. Distribution of modes,
Ann. Phys. (New York) {\bf 104}, 300 (1977).

\bibitem{Nesterenko}
V. V. Nesterenko,
Simple method for calculating the Casimir energy for a sphere,
Phys. Rev. D {\bf 57}, 1284 (1998).

\bibitem{LeoBook}
U. Leonhardt, 
{\it Essential Quantum Optics: From Quantum Measurements to Black Holes}, 
(Cambridge University Press, Cambridge, 2010).

\bibitem{Divergence}
Relation (\ref{div}) follows from Eq.~(\ref{tensordiv}) in spherical coordinates with $ds^2=dr^2+r^2d\theta^2+r^2\sin^2\theta\, d\phi^2$.

\bibitem{Survey}
J. R. Gott III, M. Juri\'{c}, D. Schlegel, F. Hoyle, M. Vogeley, M. Tegmark, N. Bahcall, and J. Brinkmann,
A Map of the Universe,
Astrophys. J. {\bf 624}, 463 (2005).

\bibitem{Fish3D}
U. Leonhardt and T. G. Philbin, 
Perfect imaging with positive refraction in three dimensions, 
Phys. Rev. A {\bf 81}, 011804 (2010).

\bibitem{GH}
U. Leonhardt,
Cosmological horizons radiate,
Europhys. Lett. {\bf 135}, 10002 (2021).

\bibitem{DiValentino}
E. Di Valentino, O. Mena, S. Pan, L. Visinelli, W. Yang, A. Melchiorri, D. F. Mota, A. G. Riess, and J. Silk,
In the Realm of the Hubble tension --- a Review of Solutions,
Class. Quantum Grav. {\bf 38}, 153001 (2021).

\bibitem{Bjorkholm}
J. E. Bjorkholm, R. H. Freeman, A. Ashkin, and D. B. Pearson,
Observation of Focusing of Neutral Atoms by the Dipole Forces of Resonance-Radiation Pressure,
Phys. Rev. Lett. {\bf 41}, 1361 (1978).

\bibitem{Spectroscopy}
R. Ozeri, N. Katz, J. Steinhauer, and N. Davidson,
Colloquium: Bulk Bogoliubov excitations in a Bose-Einstein condensate,
Rev. Mod. Phys. {\bf 77}, 187 (2005).

\bibitem{Erdelyi}
A. Erd\'{e}lyi, W. Magnus, F. Oberhettinger, and F. G. Tricomi,
{\it Higher Transcendental Functions}
(McGraw-Hill, New York, 1981).

\bibitem{Bres}
C.-S. Br\'{e}s, 
With a fine-tooth comb,
Nat. Phys. {\bf 16}, 600 (2020).

\bibitem{Barrera}
R. G. Barrera, G. A. Estevez, and J. Giraldo,
Vector spherical harmonics and their application to magnetostatics,
Eur. J. Phys. {\bf 6}, 287 (1985).

\bibitem{MagRemark}
One should use the homogeneous wave equation here, {\it i.e.} Eq.~(\ref{eq:G}) with right--hand side equal to zero, because $\bm{r}_0\neq\bm{r}$ (before the limit $\bm{r}_0\rightarrow\bm{r}$ is taken).

\bibitem{Langer}
R. E. Langer,
On the connection formulas and the solutions of the wave equation,
Phys. Rev. {\bf  51}, 669 (1937).

\bibitem{Olver}
F. Olver,
{\it Asymptotics and special functions},
(CRC Press, London, 1997).


\end{thebibliography}
\end{document}